\documentclass[a4paper,12pt]{article}
\usepackage{eurosym}
\usepackage[bookmarks, linkcolor=blue, urlcolor=blue]{hyperref}
\usepackage{xr-hyper}
\usepackage{amsfonts}
\usepackage{amssymb}
\usepackage{amsthm}
\usepackage{amsmath}
\usepackage{graphicx}
\usepackage{rotating}
\usepackage{lscape}
\usepackage{subfig}
\usepackage[font=small]{caption}
\usepackage{float}
\usepackage{lscape}
\usepackage{bm}
\usepackage{comment}
\usepackage{placeins}
\usepackage{bm}
\usepackage{mathtools}
\usepackage{setspace}
\usepackage{hyperref}
\usepackage{graphicx}
\usepackage{xcolor,soul}
\usepackage{adjustbox}
\usepackage[authoryear]{natbib}
\usepackage{dsfont}
\usepackage[title]{appendix}

\specialcomment{mycomment} {\begingroup\color{gray}}{\endgroup}
\excludecomment{mycomment}
\definecolor{MyDarkBlue}{rgb}{0.1,0.2,0.65}
\hypersetup{
colorlinks   = true,
citecolor    = MyDarkBlue
}
\newtheorem{theorem}{Theorem}

\newtheorem{assumption}{Assumption}
\newtheorem{assumptionalt}{Assumption}[assumption]
\newenvironment{assumptionstar}[1]{
\renewcommand\theassumptionalt{#1}
\assumptionalt
}{\endassumptionalt}
\newtheorem*{ass*}{Assumption $\boldsymbol{7^{\ast}'}$}
\newtheorem*{ass2*}{Assumption $\boldsymbol{2''}$}

\newenvironment{pff}[1][Proof]{\vspace{1ex}{\noindent\textbf{#1.} }\hspace{.1em}}
{\hfill\qed\vspace{1ex}}
\numberwithin{equation}{section}
\newdimen\origiwspc
\newdimen\origiwstr

\begin{document}

\title{\textbf{Nonparametric \textquotedblleft rich
covariates\textquotedblright \vspace{-12pt}\\
without saturation}\thanks{\setlength{\baselineskip}{1.2em}Acknowledgements: We are grateful to Julian Costas-Fernandez, Myoung-jae Lee, Paulo Parente, Matthias Parey, Myungkou Shin, Stefan Sperlich, and participants at the 2025 edition of the Bristol Econometrics Study Group Conference for helpful discussions, comments, and suggestions. The usual disclaimer applies. Glórias gratefully acknowledges funding by Fundação Para a Ciência e a Tecnologia: Bolsa de Investigação para Doutoramento 2023.01622.BD.}\vspace{-6pt}}

\date{23 July 2025 }
\author{\setlength{\baselineskip}{1.2em}Ludgero Gl\'orias\thanks{\setlength{\baselineskip}{1.2em}
School of Social Sciences, University of Surrey, United Kingdom. \textit{E-mail:} lg01113@surrey.ac.uk}
\and
Federico Martellosio\thanks{\setlength{\baselineskip}{1.2em}
School of Social Sciences, University of Surrey, United Kingdom. \textit{E-mail:} f.martellosio@surrey.ac.uk}
\and
J.M.C. Santos Silva\thanks{\setlength{\baselineskip}{1.2em}
School of Social Sciences, University of Surrey, United Kingdom. \textit{E-mail:} jmcss@surrey.ac.uk}}
\maketitle

\begin{abstract}
\noindent We consider two nonparametric approaches to ensure that linear instrumental variables estimators satisfy the
rich-covariates condition emphasized by \cite{blandhol2025}, even when the instrument is not unconditionally randomly assigned and the model is not saturated. Both approaches start with a nonparametric estimate of the expectation of the instrument conditional on the covariates, and ensure that the rich-covariates condition is satisfied either by using as the instrument the difference between the original instrument and its estimated conditional expectation, or by adding the estimated conditional expectation to the set of regressors. We derive asymptotic properties when the first step uses kernel regression, and assess finite-sample performance in simulations where we also use neural networks in the first step. Finally, we present an empirical illustration that highlights some significant advantages of the proposed methods.\newline
\end{abstract}

\vspace{.4cm}
\noindent \emph{JEL codes}: C14, C26, C45.

\noindent \emph{Key words}: Average treatment effect; Instrumental
variables; Local average treatment effect; Semiparametric estimation.\medskip
\pagebreak

\section{Introduction}

Linear instrumental variables estimators (LIVEs) have become one of the workhorses of the program evaluation literature, being regularly used to estimate the causal effect of policies with heterogeneous effects (see, e.g., \citealp{angrist2009}, and \citealp{angrist2010}). Indeed, based on the pioneering work of \cite{Imbens1994} and \cite{angrist1995}, LIVEs are generally viewed as identifying the average treatment effect for the sub-population of compliers (the local average treatment effect or LATE), or a weighted average of LATEs; see, e.g., \cite{Imbens1994}, \cite{angrist1995}, \cite{angrist2000}, and \cite{mogstad2021}.

However, several authors have recently pointed out that, in the standard case in which the model includes covariates, the causal interpretation of estimates obtained using LIVEs requires some strong assumptions. In particular, building on the work of \cite{abadie2003} and \cite{kolesar2013}, \cite{blandhol2025} emphasized that, in a just-identified model with a single endogenous explanatory variable and covariates, LIVEs only identify a parameter with a causal interpretation if the expectation of the instrument conditional on the covariates is linear. In the words of \cite{blandhol2022}, when this condition is satisfied the model is said to have rich covariates.\footnote{The introduction of covariates also has implications for the so-called
monotonicity condition; see \cite{sloczynski2024} and \cite{mogstad2024} for details. This is not the focus of the paper and we assume that the necessary monotonicity condition is satisfied.}

\citet[pp.~10--11]{kolesar2013} notes that the rich-covariates condition is uncontroversial when the model is saturated, or when the instrument is unconditionally randomly assigned, and \cite{blandhol2025} state that, outside of these two special cases, having rich covariates is a parametric assumption that needs to be defended \citep[see also][]{mogstad2024}.

In this paper, we propose methods to ensure that the rich-covariates condition is satisfied without imposing parametric assumptions, and without relying on model saturation or randomly assigned instruments. The estimators we propose make use of a preliminary estimate of the expectation of the instrument conditional on the covariates, which can be obtained nonparametrically. We then consider two ways of using this estimate to ensure that the rich-covariates condition is fulfilled, at least asymptotically, even in cases where the instrument is not unconditionally randomly assigned and the model is not saturated. The two approaches we consider are closely related to the two cases that \cite{kolesar2013} identified as ensuring rich covariates.

The first method, which we refer to as the instrument-residual approach following \cite{lee2021}, uses as the instrument the difference between the original instrument and its estimated conditional expectation. This ensures that the rich-covariates condition is satisfied because the difference between the instrument and its conditional expectation is mean independent of all functions of the covariates, and therefore its conditional expectation is constant (more specifically zero). The second method is a control-function approach in which the estimated conditional expectation of the instrument is added to the set of explanatory variables. This ensures that the rich-covariates condition is satisfied because the conditional expectation of the instrument is trivially a linear function of itself.

The estimators we propose have as special cases other estimators considered in the literature. In particular, if the conditional expectation of the instrument is estimated using a saturated linear model, the two methods we propose lead to results that are numerically identical to those obtained by estimating the saturated model discussed, for example, by \cite{angrist1995}, \cite{kolesar2013}, \cite{mogstad2024}, and \cite{blandhol2025}. The advantage of our approach is that other nonparametric estimators of the conditional expectation can be used, which is particularly convenient when the set of controls includes continuous covariates or discrete covariates with large support. Naturally, like saturation, nonparametric estimators of the conditional expectation may be affected by the curse of dimensionality, and therefore implementation may be challenging in models with many controls; we discuss this issue in Section \ref{sec approaches}.

In turn, our instrument-residual estimator has as a special case the estimators proposed by \cite{lee2021} and \cite{kim2024}; see also \cite{lee2024}. Our approach differs from that of Lee and co-authors in two important ways. First, we argue that the parametric approaches used by these authors to estimate the conditional expectation of the instrument are far too restrictive, and emphasize the importance of using nonparametric estimators. Second, our approach accommodates an extensive set of functions of the control variables in the estimating equation, potentially leading to substantial efficiency gains.

The instrument-residual estimator is also closely related to the double/debiased machine learning (DML) estimators of the partially linear instrumental variables model considered by \cite{chernozhukov2018}, which also use an instrument residual.\footnote{\cite{LeeandLee25} discuss estimators closely related to  those suggested by \cite{chernozhukov2018}.} Their estimators require the nonparametric estimation of several different objects and involve a computationally-expensive cross-fitting step, therefore being more difficult to implement than our estimators. In Section \ref{sec approaches} we provide further details on the relation between our proposed instrument-residual estimator and those suggested by \cite{chernozhukov2018}.

Crucially, under standard conditions, both of our estimators identify a positively-weighted average of expected treatment effects for compliers, conditional on the set of covariates. Therefore, the parameter that our estimators\textbf{\ } identify is weakly causal, in the sense of \cite{blandhol2025}, and has a clear interpretation. Moreover, this estimand is the one identified by the LIVE under rich covariates and coincides with the estimand of the estimators proposed by \cite{lee2021} and \cite{kim2024}, and with the estimand of the DML estimator when there are heterogeneous treatment effects.  

In short, the estimators we propose are in between the instrument-residual estimators of Lee and co-authors, which are implemented using strong parametric assumptions and can be very inefficient, and the DML estimators of Chernozhukov and co-authors, which require the nonparametric estimation of more objects and rely on cross-fitting which, as we will show, brings its own problems. 

Although with a very different motivation, the two approaches we suggest parallel the ones considered by \cite{BH23} in a related but distinct context.\footnote{We are grateful to Julian Costas-Fernandez for pointing to us the parallel between our approaches and those of \cite{BH23}.} In contrast to what happens in \cite{BH23}, we cannot use simulation methods to estimate the expectation of the instrument conditional on the covariates, and have to rely on other non-parametric estimators such as kernel regression or machine-learning methods.

The use of a nonparametric estimate of the conditional expectation of the instrument raises the possibility that the estimate of the parameter of interest may not be $\sqrt{n}$-consistent. However, it is well known that estimators can converge at the parametric rate even when they depend on a preliminary nonparametric step that does not converge at the same rate \citep[see, e.g.,][]{NMcF1994}. In Section \ref{sec asy} we show that, under suitable conditions, a similar result holds in our case when the nonparametric step is performed using kernel methods. It must be emphasized that this result is presented to show that our approaches can lead to estimators that are $\sqrt{n}$-consistent, and does not imply that we recommend the use of kernel regression in the preliminary nonparametric step. 

The remainder of the paper is organized as follows. Section \ref{sec setup} presents the setup and notation. Section \ref{sec approaches} introduces our proposed estimators under the assumption that the expectation of the instrument conditional on the covariates is known, and then considers the properties of the estimators when the expectation of the instrument conditional on the covariates needs to be estimated. In Section \ref{sec asy}
we show that the estimators of the parameter of interest can be $\sqrt{n}$ -consistent when the expectation of the instrument conditional on the covariates is estimated by a suitable kernel regression. Section \ref{sec simul} presents a small simulation study illustrating the performance of the proposed methods and comparing them with alternative approaches. Section \ref{sec examples} revisits one of the empirical studies considered by \cite{blandhol2025} to illustrate the application of the methods we propose and highlight their attractiveness. Finally, Section \ref{sec concl} concludes and the appendices contain further technical material and the proofs.

\section{\label{sec setup}Setup and notation}

As, for example, in \cite{kolesar2013} and \cite{blandhol2025}, we consider the case where we want
to learn about the heterogeneous causal effect of a binary treatment $t\in \left\{ 0,1\right\} $ on the outcome $y$. As usual, $y$ can be written as $y(1)t + y(0)\left( 1-t\right)$, where $y(1)$ and $y(0)$ denote the outcome with and without treatment, respectively. The treatment is potentially endogenous, but we assume that we also observe an instrument $z\in $ $\left\{ 0,1\right\} $ that is exogenous, at least when conditioning on a set of $d$ observable covariates collected in the vector $\boldsymbol{c}$. The treatment status depends on $z$, and we use $t(0)$ and $t(1)$ to denote the potential values of the treatment when $z$ equals $0$ or $1$, respectively.

In this framework, we are interested in the conditions under which a researcher can obtain information on the causal effect of $t$ on $y$ by using instrumental variables to estimate a linear equation of the form
\begin{equation}
y=\alpha t+\boldsymbol{r}^{\prime }\boldsymbol{\gamma }+\varepsilon =%
\boldsymbol{x}^{\prime }\boldsymbol{\beta }+\varepsilon ,  \label{Model}
\end{equation}
where $\boldsymbol{r}$ is a $k\times 1$ vector containing functions of $\boldsymbol{c}$ ($k\geq 0$), $\boldsymbol{x}\coloneqq (t,\boldsymbol{r}^{\prime })^{\prime }$, $\boldsymbol{\beta }\coloneqq (\alpha ,\boldsymbol{\gamma }^{\prime })^{\prime }$ is the corresponding $\left( k+1\right) \times 1$ vector of parameters, and $\varepsilon $ is an error term. We note that we focus on the LIVE of (\ref{Model}) because this is the approach commonly used by practitioners who want to estimate causal relationships, and not because we assume that such linear equation is an adequate representation of the data generating process.  For example, we do not assume that the effect of $t$ on $y$ is constant, or that there is linearity of $y$ in $r$.\footnote{It is worth noting, however, that linearity of $y$ in $r$ is used in the moment condition for the LIVE, so we would expect the estimator to be more precise if linearity in $r$ is a good approximation. }

For the instrumental variable estimate of $\alpha $ to have standard properties and a causal interpretation, the instrument needs to satisfy a number of conditions. In particular, the instrument needs to be exogenous, in the sense that $\left( y\left( 0\right) ,y\left( 1\right) ,t\left(0\right) ,t\left( 1\right) \right) \mathrel{\perp\!\!\!\!\perp}z\mid\boldsymbol{c}$, relevant, in the sense that $\mathbb{C}\text{ov}( z,t\mid\boldsymbol{c}) \neq 0$, and a suitable form of monotonicity needs to hold; see \cite{sloczynski2024} and \cite{mogstad2024} for details.\footnote{Specifically, as we will see in Section \ref{sec approaches}, we need that the sign of $\mathbb{C}\text{ov}( z,t\mid\boldsymbol{c}) $ is the same for all $\boldsymbol{c}$. The validity of this assumption has to be considered on a case-by-case basis.} Here, we assume that those other necessary conditions hold and, as in \cite{blandhol2025}, focus on the rich-covariates condition, which requires that 
\begin{equation}
\mathbb{E}\left[ z\mid\boldsymbol{c}\right] =\mathbb{L}\left[ z\mid\boldsymbol{r}%
\right] ,  \label{RC}
\end{equation}
where $\mathbb{L}\left[ z\mid\boldsymbol{r}\right] $ denotes the linear projection of $z$ on $\boldsymbol{r}$.\footnote{\cite{blandhol2025} write the rich-covariates condition as $\mathbb{E}%
\left[ z\mid\boldsymbol{c}\right] =\mathbb{L}\left[ z\mid\boldsymbol{c}\right] $ but it is useful to distinguish the set of conditioning variables from the set of functions of those variables that are included in the estimating equation.} \citet[p.~10]{kolesar2013} calls (\ref{RC}) the linearity assumption, but we prefer to use the rich-covariates label introduced by \cite{blandhol2022} because linearity can be misconstrued to relate to the linearity of (\ref{Model}), to the linearity of the data generation process, or to the linearity of the conditional expectation $\mathbb{E}\left[y(t)\mid\boldsymbol{c}\right] $ for $t\in $ $\left\{ 0,1\right\} $.

\section{\label{sec approaches}The proposed approaches}

This section introduces the two approaches we propose. As, for example, in \cite{RMN92} and \cite{abadie2003}, we start by considering the infeasible \textquotedblleft oracle\textquotedblright\ estimators obtained assuming that $\zeta _{0}(\boldsymbol{c})\coloneqq\mathbb{E}[z\mid\boldsymbol{c}]$ is known, and then we consider the implications of replacing the unknown conditional expectation by its estimated counterpart.

\subsection{The instrument-residual approach}

As mentioned before, Koles\'{a}r (2013) noted that the rich-covariates condition is uncontroversial when the instrument is unconditionally randomly assigned. In this case the instrument is statistically independent of $\boldsymbol{c}$, and consequently its conditional expectation is constant. However, the instrument does not need to be statistically independent of the controls for its conditional expectation to be constant, it is enough for the instrument to be mean independent of $\boldsymbol{c}$. Therefore, given $z$ and $\zeta _{0}( \boldsymbol{c}) $, we can ensure that the model has rich covariates by using the \textit{instrument residual} $z^{\ast}( \zeta _{0}) \coloneqq z-\zeta _{0}( \boldsymbol{c}) 
$ as the instrument for $t$.

We are not the first to propose an instrument-residual approach. Among others, \cite{lee2021} and \cite{kim2024} have also proposed estimators that use $z^{\ast }( \zeta _{0}) $ as the instrument  \citep[see also][]{BH21}. Their estimators can be interpreted as the LIVE of $y$ on $t$ using the instrument residual\ $z^{\ast }( \zeta _{0}) $, in an estimating equation that has no additional controls. For the case where $z$ is binary, Lee (2021) shows that his estimator identifies
\begin{equation}
\alpha _{\text{rich}}\coloneqq \frac{\mathbb{C}\text{ov}( z^{\ast }(
\zeta _{0}) ,y) }{\mathbb{C}\text{ov}( z^{\ast }( \zeta
_{0}) ,t) }=\mathbb{E}\left[ \omega ( \boldsymbol{c}) 
\mathbb{E}_{\text{C}}( y(1)-y(0)\mid\boldsymbol{c}) \right],  \label{alpharich}
\end{equation}
where, as in \cite{blandhol2025}, the subscript ``rich'' indicates that this is the parameter identified by LIVEs when the rich-covariates condition is satisfied, $\mathbb{E}_{\text{C}}$ denotes that expectations are taken over the sub-population of compliers, and $\omega ( \boldsymbol{c}) $
are the so-called overlap weights defined as
\begin{equation*}
\omega ( \boldsymbol{c}) \coloneqq\frac{\mathbb{C}\text{ov}( z,t\mid%
\boldsymbol{c}) }{\mathbb{E}\left[ \mathbb{C}\text{ov}( z,t\mid\boldsymbol{c}) %
\right] },
\end{equation*}%
with $\mathbb{C}\text{ov}( z,t\mid\boldsymbol{c}) =\Pr \left( \text{Complier}\mid\boldsymbol{c}\right) \zeta _{0}( \boldsymbol{c}) \left( 1-\zeta_{0}( \boldsymbol{c}) \right) $, where $\Pr \left( \text{Complier}\mid\boldsymbol{c}\right) $ denotes the conditional probability of being a complier.\footnote{\citet{sloczynski2020,sloczynski2024} also presents this result.} That is, the estimator gives additional weight to individuals who are more likely to comply, and therefore for whom the instrument is stronger, and to individuals for whom $\boldsymbol{c}$ has a weaker relation with the instrument, i.e., for whom the instrument is closer to being randomly allocated \citep[see][and the references therein]{lee2021}. 

Equation (\ref{alpharich}) shows that $\alpha _{\text{rich}}$ is weakly causal, in the sense of \cite{blandhol2025}, as long as the sign of $\mathbb{C}\text{ov}( z,t\mid\boldsymbol{c}) $ is the same for all $\boldsymbol{c}$; that is, as long as the strong monotonicity condition holds \citep[see, e.g.,][]{sloczynski2024, mogstad2024}. \cite{kim2024} show that this result can be generalized to the case where the instrument is multi-valued, and note that continuous instruments can be discretized. \cite{BH21} address a closely related problem and present the estimand for the corresponding instrumental-residual estimator when the instrument is continuous. Crucially, all these results do not depend on (\ref{Model}) being correctly specified in any reasonable sense.

It is easy to see that adding covariates to the model does not change the probability limit of the estimator for the coefficient on $t$ (see \citealp{BH21}, and Appendix \ref{App estimands}). Therefore, our instrument-residual estimator has the same estimand as the estimators of \cite{lee2021} and \cite{kim2024}, but our estimator can be much more efficient because the additional variables included in $\boldsymbol{r}$ reduce the variance of the error. Indeed, controlling for additional covariates is often motivated as a way to gain efficiency (see, e.g., \citealp{abadie2003} and \citealp{BH23}).

\cite{kim2024} suggest the parametric estimation of $\zeta _{0}\left( \boldsymbol{c}\right) $, but note that a nonparametric alternative is to use the DML estimators of the partially linear instrumental variables model presented by \cite{chernozhukov2018}; \cite{blandhol2025} also consider this estimator as an alternative to the LIVE.

The estimators introduced by \cite{chernozhukov2018} are closely related to our instrument-residual estimator in that they also use an instrument residual. Indeed, \citet[pp.~C32--C34]{chernozhukov2018} consider the estimation of a partially linear instrumental variables model of the form
\begin{equation*}
y=\rho t+\xi ( \boldsymbol{c}) +v,
\end{equation*}
where $\xi ( \boldsymbol{c}) $ is a function of $\boldsymbol{c}$ and $\mathbb{E}\left[ v\mid\boldsymbol{c},z\right] =0$.  The authors consider estimators for $\rho $, the parameter of interest, based on two alternative moment conditions, both of which use $z^{\ast }( \zeta _{0}) $ as the instrument. Their preferred
approach is based on the score (see their equation 4.8)
\begin{equation*}
\big( y-\mathbb{E}\left[ y\mid\boldsymbol{c}\right] -\rho \left( t-\mathbb{E}%
\left[ t\mid\boldsymbol{c}\right] \right) \big) z^{\ast }( \zeta
_{0}) ,
\end{equation*}
where, besides $\zeta _{0}( \boldsymbol{c}) $, $\mathbb{E}\left[ y\mid\boldsymbol{c}\right] $ and $\mathbb{E}\left[ t\mid\boldsymbol{c}\right] $ have to be estimated nonparametrically. The alternative estimator is based on the score (see their equation 4.7)
\begin{equation*}
\left( y-\rho t-\xi ( \boldsymbol{c}) \right) z^{\ast }(\zeta _{0}) ,
\end{equation*}
which is more difficult to operationalize because $\xi \left( \boldsymbol{c} \right) $ cannot be directly estimated.\footnote{\citet[fn.~8]{chernozhukov2018} also consider a different score, but that is not based on $z^{\ast }( \zeta _{0}) $.}

Both of these scores can be expressed as
\begin{equation*}
\left( y-\rho t-m( \boldsymbol{c}) \right) z^{\ast }( \zeta
_{0}) ,
\end{equation*}
where $m( \boldsymbol{c}) $ is a function of $\boldsymbol{c}$. However, because the instrument residual $z^{\ast }( \zeta _{0}) $ is mean independent of any function of $\boldsymbol{c}$, the form of $m( \boldsymbol{c}) $ does not affect the estimand, but it affects the efficiency of the estimator.\footnote{Indeed, $m( \boldsymbol{c}) $ can even be omitted from the score function, as in \cite{lee2021} and \cite{kim2024}.}

Therefore, our estimator can be seen as a simplification of the estimators proposed by \cite{chernozhukov2018}, in which $m( \boldsymbol{c}) $ is specified by the practitioner as $\boldsymbol{r}^{\prime } \boldsymbol{\gamma }$.  Naturally, \citeauthor{chernozhukov2018}'s (\citeyear{chernozhukov2018}) approach has the advantage of controlling for $\boldsymbol{c}$ nonparametrically, and that will often lead to increased efficiency, at the cost of being computationally more expensive.\footnote{Additionally, because it relies on estimates of $\mathbb{E}\left[ y\mid\boldsymbol{c}\right] $ and $\mathbb{E}\left[ t\mid\boldsymbol{c}\right] $,  \citeauthor{chernozhukov2018}'s (\citeyear{chernozhukov2018}) estimator requires assumptions about these functions, which we do not generally need.} 

A key feature of the DML estimators introduced by \cite{chernozhukov2018} is the fact that they are based on Neyman-orthogonal scores, which reduces the sensitivity of the estimator of $\rho $ to the noise in the estimates of the nonparametric functions \citep[see also][]{LeeandLee25}. We show in Appendix \ref{App Neyman} that the score on which our instrument-residual estimator is based would satisfy the Neyman-orthogonality condition if and only if (\ref{Model}) were correctly specified, in the sense that $\mathbb{E}\left[ \varepsilon \mid \boldsymbol{c}\right] =0$, which is an assumption we do not make.\footnote{Note, however, that because the score implied by our estimator nests the one of \citeauthor{chernozhukov2018}'s (\citeyear{chernozhukov2018}) preferred estimator, the condition  $\mathbb{E}\left[ \varepsilon \mid \boldsymbol{c}\right] =0$ would be satisfied if $\boldsymbol{r}$ were to include estimates of $\mathbb{E}\left[ y\mid\boldsymbol{c}\right] $ and $\mathbb{E}\left[ t\mid\boldsymbol{c}\right] $, in which case the score on which our estimator is based would be Neyman orthogonal.} Because, in general, the score on which our estimator is based is not Neyman orthogonal, we cannot use the results in \cite{chernozhukov2018} to establish its properties. As a consequence, our estimator does not use cross-fitting which, besides being computationally expensive, will be shown to be unsuitable in some empirically-relevant scenarios.

\subsection{The control-function approach}

The second situation identified by \cite{kolesar2013} where the rich-covariates condition is uncontroversial is in saturated models. This shows that the fulfillment of the rich-covariates condition does not depend on $\boldsymbol{c}$ but on how $\boldsymbol{c}$ is used to specify the equation used to estimate the causal effect of $t$ on $y$. We consider an alternative specification to equation (\ref{Model}) that ensures the model has rich covariates.

Observing that
\begin{equation*}
\mathbb{L}\left[ z\mid\boldsymbol{r},\zeta _{0}( \boldsymbol{c}) %
\right] =\zeta _{0}( \boldsymbol{c}) ,
\end{equation*}
it follows that the rich-covariates condition is satisfied as long as (\ref{Model}) is augmented by the addition of $\zeta \left( \boldsymbol{c} \right) $ as an explanatory variable. That is, the estimate of interest is obtained by applying a LIVE to the extended equation 
\begin{equation}
y=\alpha ^{\ast }t+\boldsymbol{r}^{\prime }\boldsymbol{\gamma }^{\ast }+\phi
\zeta _{0}(\boldsymbol{c})+\varepsilon ^{\ast }=\boldsymbol{x}^{\ast \prime }%
\boldsymbol{\beta }^{\ast }+\varepsilon ^{\ast },  \label{ExM}
\end{equation}
where $\boldsymbol{x}^{\ast }\coloneqq(t,\boldsymbol{r}^{\prime },\zeta _{0}( \boldsymbol{c}))^{\prime }$ is a $\left( k+2\right) \times 1$ vector, $ \boldsymbol{\beta }^{\ast }\coloneqq(\alpha ^{\ast },\boldsymbol{\gamma }^{\ast \prime },\phi )^{\prime }$ is the corresponding $\left( k+2\right)
\times 1$ vector of unknown parameters, $\varepsilon ^{\ast }$ in an error term, and the other symbols are defined as before. As in the instrument-residual approach, equation (\ref{ExM}) defines the moment condition upon which our estimator is based, but is not assumed to be correctly specified.

In Appendix \ref{App estimands} we show that the estimators of $\alpha^{\ast} $ and $\gamma^{\ast}$ using $z$ as an instrument for $t$ yield the same estimands as the instrument-residual approach \cite[see also][]{BH23}. As before, it is helpful if $\boldsymbol{r}$ includes an extensive set of functions of $ \boldsymbol{c}$ because that has the potential to improve the efficiency of the estimator. Additionally, we show in Appendix \ref{App Neyman} that the score on which estimation of $\alpha $ is based is Neyman orthogonal only under very strong conditions that generally do not hold.\footnote{For example, the score will be Neyman-Orthogonal when the pseudo-true value of $\phi$ is $0$ and $\mathbb{E}\left[ \varepsilon ^{\ast}\mid\boldsymbol{c}\right] =0$. Once again, these conditions hold when $\boldsymbol{r}$ is composed of estimates of $\mathbb{E}\left[ y\mid\boldsymbol{c}\right] $ and $\mathbb{E}\left[ t\mid\boldsymbol{c}\right] $.} 

Given that the two proposed approaches identify the same parameter, it is natural to ask which of them is preferable in practice. We will address this question later, but for now we note that, in contrast to what happens in the
instrument-residual approach, in the control-function approach the instrument is binary, which may be relevant in some circumstances. For example, the control-function approach can be used to ensure rich covariates when using the $\kappa $-weighted regression suggested by \cite{abadie2003}; see also \citet[Section 5.3]{blandhol2025}.

\subsection{Allowing for the estimation of the conditional expectation}

We now discuss the consequences of using different estimators of $\zeta _{0}(\boldsymbol{c})$, consider the effects of this choice on the properties of the estimators of $\alpha $, and motivate the estimators used in the
simulation exercise performed in Section \ref{sec simul} and in the empirical illustration presented in Section \ref{sec examples}. However, we emphasize that the optimal choice of the estimator of $\zeta _{0}(\boldsymbol{c})$ may vary from application to application and that future developments in nonparametric estimation may lead to different recommendations on the choice of method to use in the preliminary step.

\cite{lee2021} and \cite{kim2024} suggest estimating $\zeta _{0}(\boldsymbol{c})$ by parametric methods, and that is the approach implemented in \cite{lee2024}. The obvious advantages of this approach are its simplicity and the fact that, as in the classic works by \cite{pagan1984} and \cite{murphy2002}, the estimators of $\alpha $ will converge at the usual $\sqrt{n}$ rate (\citeauthor{RMN92}, \citeyear{RMN92}, and \citeauthor{abadie2003}, \citeyear{abadie2003}, for example, also consider parametric approaches in related contexts). The drawback, however, is that misspecification of the parametric model for $\zeta _{0}(\boldsymbol{c})$ implies that the rich-covariates condition is not fulfilled and there is no guarantee that the estimator of $\alpha $ can be given a causal interpretation. Therefore, except possibly in applications where economic theory provides some guidance on its form, we find it difficult to recommend the use of a parametric estimator of $\zeta _{0}(\boldsymbol{c})$.

The additional robustness afforded by using a nonparametric method to estimate $\zeta _{0}(\boldsymbol{c})$ has non-negligible costs. A first potential drawback is that the resulting estimator of $\alpha $ may not converge at the parametric rate. However, it is well known that under suitable regularity conditions, estimators can be $\sqrt{n}$-consistent even when they depend on a preliminary nonparametric step that does not converge at the same rate (see, e.g., \citealp{robinson1988}, \citeauthor{newey1994}, \citeyear{newey1994}, \citeyear{newey1997}, \citealp{NMcF1994}, and \citealp{newey2004}). For example, \cite{abadie2003} obtains an estimate of $\zeta _{0}(\boldsymbol{c})$ using power series estimators and, using the results of \cite{newey1994}, shows that the second step estimator is $\sqrt{n}$-consistent. In the next section we show that, under suitable conditions, a similar result holds in our case when $\zeta _{0}(\boldsymbol{c})$ is estimated using kernel methods. In the simulations in Section \ref{sec simul} we illustrate the performance of our estimators when the preliminary step is performed in this way.

A related cost of using a nonparametric preliminary step is the well-known curse of dimensionality. In empirical applications, $d$ (the dimension of $\boldsymbol{c}$) is often large and that may severely affect the performance of series and kernel estimators. A possible alternative is to estimate $\zeta _{0}(\boldsymbol{c})$ using machine learning methods such as neural networks, which \cite{barron1994} showed can be less affected by the curse of dimensionality than series- and kernel-based estimators. Indeed, recent
results suggest that neural network estimators may be able to circumvent the curse of dimensionality under certain conditions \citep[see, e.g.,][]{bach2017,bauer2019,schmidt2020,kohler2021,braun2024}. Neural networks have the additional practical advantage of being able to seamlessly handle both continuous and discrete controls.

The properties of neural networks with different architectures is currently a very active area of research, so it is not yet clear what would be the optimal way to obtain a preliminary estimate of $\zeta _{0}(\boldsymbol{c})$ based on such methods. In the simulations in Section \ref{sec simul}, we study the performance of our estimators when $\zeta _{0}(\boldsymbol{c}) $ is estimated using a neural network based on the rectified linear unit (ReLU) activation function, with a single hidden layer with $100$ nodes, as implemented with the default options in the user-written \texttt{pystacked} Stata command \citep{ahrens2023}.\footnote{\texttt{pystacked} uses scikit-learn's machine learning algorithms, see \cite{scikit-learn}.} This kind of neural network shares some of the features of the ones considered by \cite{bach2017}, and therefore it is likely to have a good performance \citep[see also,][]{braun2024}.\footnote{Alternatively, we could have used deep networks, such as those considered by \cite{schmidt2020}, \cite{kohler2021}, and \cite{farrell2021}, which also have attractive properties. However, the results in \cite{Hornik1989} suggest that the number of hidden nodes is more important than the number of layers. Comparing the performance of alternative network architectures is left for future work.} 

An additional advantage of using neural network estimators is that, in contrast with lasso, random forests, and other popular machine learning methods, they do not rely heavily on regularization \citep[see, e.g.,][]{farrell2021}. Specifically, by default, the neural network estimator implemented in \texttt{pystacked} uses $L_2$ regularization with a very small penalty parameter, which is then divided by the sample size. Therefore, these estimates will have minimal regularization bias. This is specially important for the methods that we propose because they are based on moment conditions that generally are not Neyman-orthogonal.

Finally, we note that, just like in the case of series and kernel estimators, it is possible to obtain $\sqrt{n}$-consistent estimators for the parametric part of semiparametric models when neural networks or other machine learning methods are used in a preliminary nonparametric step \citep[see, e.g.,][]{chen1999,chernozhukov2018, farrell2021}. Studying the conditions under which it is possible to obtain a $\sqrt{n}$-consistent estimator of the causal effect of the treatment when $\zeta _{0}(\boldsymbol{c})$ is estimated using a neural network is, however, left for future work.

\section{\label{sec asy}Asymptotics with kernel first step}

In this section we show that, under the set of assumptions detailed below,
the LIVEs of the causal effect of the treatment in equations (\ref{Model})
and (\ref{ExM}) are $\sqrt{n}$-consistent and asymptotically normal when $%
\zeta _{0}(\boldsymbol{c})$ is estimated using a suitable kernel regression
estimator. The main device we use to achieve the parametric rate for the
LIVEs, despite the fact that the first-step kernel estimator has slower rate
of convergence, is undersmoothing \citep[see][]{NMcF1994}.

\subsection{First-step kernel estimation}

Based on a random sample of size $n$, the conditional expectation $\mathbb{E}%
\left( z\mid\boldsymbol{c}\right) =\zeta _{0}(\boldsymbol{c})$ can be estimated
by the Nadaraya-Watson estimator \citep[][]{nadaraya1964,watson1964}
\begin{equation}
\hat{\zeta}( \boldsymbol{c}) \coloneqq\frac{\sum_{i=1}^{n}K_{h}%
\left( \boldsymbol{c}-\boldsymbol{c}_{i}\right) z_{i}}{\sum_{i=1}^{n}K_{h}%
\left( \boldsymbol{c}-\boldsymbol{c}_{i}\right) },  \label{NW}
\end{equation}%
where $K_{h}(\boldsymbol{u})\coloneqq h^{-d}K(h^{-1}\boldsymbol{u})$, with $%
K(\boldsymbol{u})$ being a kernel whose properties are specified below, and $%
h=h(n)$ is a bandwidth term. We make the following assumptions about the
kernel, the bandwidth, and the covariates.

\begin{assumption}
\label{assumption kernel}$K(\boldsymbol{u})$ is zero outside a bounded set, $%
\int K(\boldsymbol{u})\,\mathrm{d}\boldsymbol{u}=1$, and there is a positive
integer $m$ such that for all $J<m $, $\int K(\boldsymbol{u})\left(
\bigotimes_{j=1}^{J}\boldsymbol{u}\right) \,\mathrm{d}\boldsymbol{u}=%
\boldsymbol{0}_{d^{J}}$, where $\otimes $ denotes the Kronecker product of
matrices.
\end{assumption}

\begin{assumption}
\label{assumption h}The bandwidth $h$ satisfies $nh^{2d}/(\ln
n)^{2}\rightarrow \infty $ and $nh^{2m}\rightarrow 0$.
\end{assumption}

\begin{assumption}
\label{assumption c}The support $\mathcal{C}$ of $\boldsymbol{c}$ is compact
and there are $a,b>0$ such that $a<f(\boldsymbol{c})<b$ for all $\boldsymbol{%
c}$.
\end{assumption}

\begin{assumption}
\label{assumption smoothness}There is a version of $\zeta _{0}(\boldsymbol{c}%
)$ that is continuously differentiable of order $m$ and is bounded on an
open set containing $\mathcal{C}$.
\end{assumption}

Assumption \ref{assumption kernel} states that the kernel must be of order $%
m $. The first condition in Assumption \ref{assumption h} controls the
variance of $\hat{\zeta}( \boldsymbol{c}) $, while the second
condition forces undersmoothing, compared to the MSE-minimizing bandwidth
(which is proportional to $n^{-1/(2m+d)}$; see \citealp{scott2015}). Taken together, the two
conditions require that $m>d$ (so the kernel must be of higher order as long
as $d>1$) and that the bandwidth goes to $0$ at the rate $n^{-a}$, with $%
1/(2m)<a<1/(2d)$. Assumption \ref{assumption c} is standard in kernel regression analysis. It excludes discrete covariates for simplicity, but extensions of our semiparametric estimators to settings where some or all covariates are discrete are possible \citep[see, e.g.,][]{Gao2015}. Assumption \ref%
{assumption smoothness} ensures that the bias of the Nadaraya-Watson
estimator decays sufficiently fast when using a kernel of order $m$; this
level of smoothness is standard in kernel-based semiparametric estimation
and is necessary to achieve $\sqrt{n}$-consistency of the second-step
estimator under undersmoothing.

\subsection{Asymptotics for the instrument-residual approach}

We now consider equation (\ref{Model}) and define the vector $\boldsymbol{q}%
(\zeta )\coloneqq(z^{\ast }(\zeta ),\boldsymbol{r}^{\prime })^{\prime }$.
Also, let $\boldsymbol{q}_{0}\coloneqq\boldsymbol{q}(\zeta _{0})$ and, for $%
i=1,\ldots ,n$, $\boldsymbol{q}{_{i}}(\hat{\zeta})\coloneqq\bigl(z_{i}^{\ast
}(\hat{\zeta}\left( \boldsymbol{c}_{i}\right) ),\boldsymbol{r}_{i}^{\prime }%
\bigr)^{\prime }$. Based on a random sample of size $n$, the (two-step) LIVE
of $\boldsymbol{\beta }$ is%
\begin{equation*}
\boldsymbol{\hat{\beta}}\coloneqq\left( \sum_{i=1}^{n}\boldsymbol{q}{_{i}}(%
\hat{\zeta})\boldsymbol{x}_{i}^{\prime }\right) ^{-1}\left( \sum_{i=1}^{n}%
\boldsymbol{q}{_{i}(}\hat{\zeta})y_{i}\right) .
\end{equation*}

The moment condition defining $\boldsymbol{\beta }_{0}$ is%
\begin{equation}
\mathbb{E}\left[(\boldsymbol{q}_{0}(y-\boldsymbol{x}^{\prime }\boldsymbol{%
\beta }_{0})\right] =\boldsymbol{0}_{k+1},  \label{mom cond IR}
\end{equation}%
and $\boldsymbol{\hat{\beta}}$ solves the sample moment condition%
\begin{equation}
\frac{1}{n}\sum_{i=1}^{n}\left( \boldsymbol{q}_{i}(\hat{\zeta})(y_{i}-%
\boldsymbol{x}_{i}^{\prime }\boldsymbol{\beta })\right) =\boldsymbol{0}%
_{k+1}.  \label{sample mom cond IR}
\end{equation}

To proceed, we make the following additional assumptions.

\begin{assumption}
\label{assumption moments Y}$\mathbb{E}[y^{4}]<\infty $ and $\mathbb{E}%
[\left\Vert \boldsymbol{r}\right\Vert ^{4}]<\infty $.
\end{assumption}

\begin{assumption}
\label{assumption Gbeta}$\boldsymbol{G}_{\boldsymbol{\beta }}\coloneqq%
\mathbb{E}[\boldsymbol{q}_{0}\boldsymbol{x}^{\prime }]$ is non singular.
\end{assumption}

Assumption \ref{assumption moments Y} is a standard integrability condition
needed for the asymptotic analysis. Assumption \ref{assumption Gbeta} is an
identification condition guaranteeing a unique solution to moment condition (%
\ref{mom cond IR}). The following result establishes that $\boldsymbol{\hat{%
\beta}}$, the LIVE of $\boldsymbol{\beta }_{0}$, is $\sqrt{n}$-consistent
and asymptotically normal.

\begin{theorem}
\label{Th asy distr} If Assumptions \ref{assumption kernel}--\ref{assumption
Gbeta} are satisfied, then 
\begin{equation*}
\sqrt{n}\left( \boldsymbol{\hat{\beta}}-\boldsymbol{\beta }_{0}\right) 
\overset{d}{\rightarrow }\mathcal{N}\left( 0,\boldsymbol{G}_{\boldsymbol{%
\beta }}^{-1}\boldsymbol{\Omega }(\boldsymbol{G}_{\boldsymbol{\beta }%
}^{\prime })^{-1}\right) ,
\end{equation*}%
where $\boldsymbol{\Omega }\coloneqq\mathbb{V}ar\left( \boldsymbol{q}(\zeta
_{0})\varepsilon +\boldsymbol{\varphi }\right) $, with $\boldsymbol{\varphi }%
\coloneqq(-z^{\ast }(\zeta _{0})\mathbb{E}\left[ \varepsilon \mid\boldsymbol{c}%
\right] ,\boldsymbol{0}_{k}^{\prime })^{\prime }$.
\end{theorem}

The asymptotic variance $\boldsymbol{G}_{\boldsymbol{\beta }}^{-1}%
\boldsymbol{\Omega }(\boldsymbol{G}_{\boldsymbol{\beta }}^{\prime })^{-1}$
can be estimated by $\boldsymbol{\hat{G}}_{\boldsymbol{\beta }}^{-1}%
\boldsymbol{\hat{\Omega}}(\boldsymbol{\hat{G}}_{\boldsymbol{\beta }}^{\prime
})^{-1}$, where $\boldsymbol{\hat{G}}_{\boldsymbol{\beta }}\coloneqq\frac{1}{%
n}\sum_{i=1}^{n}\boldsymbol{q}_{i}(\hat{\zeta})\boldsymbol{x}_{i}^{\prime }$
and $\boldsymbol{\hat{\Omega}}\coloneqq-\frac{1}{n}\sum_{i=1}^{n}\boldsymbol{%
\hat{\tau}}_{i}\boldsymbol{\hat{\tau}}_{i}^{\prime }$, with $\boldsymbol{%
\hat{\tau}}_{i}\coloneqq\boldsymbol{q}{_{i}}(\hat{\zeta})(y_{i}-\boldsymbol{x%
}_{i}^{\prime }\boldsymbol{\hat{\beta}})+\boldsymbol{\hat{\varphi}}_{i}$,
where $\boldsymbol{\hat{\varphi}}_{i}\coloneqq(z_{i}^{\ast }(\hat{\zeta}%
\left( \boldsymbol{c}_{i}\right) )\hat{\varepsilon}_{i},\boldsymbol{0}%
_{k}^{\prime })^{\prime }$ and $\hat{\varepsilon}_{i}\coloneqq y_{i}-%
\boldsymbol{x}_{i}^{\prime }\boldsymbol{\hat{\beta}}$.\footnote{%
Note that $\boldsymbol{q}(\zeta _{0})\varepsilon +\boldsymbol{\varphi}=%
\begin{pmatrix}
z^{\ast }(\zeta _{0}) \\ 
\boldsymbol{r}%
\end{pmatrix}%
\varepsilon -%
\begin{pmatrix}
z^{\ast }(\zeta _{0}) \\ 
\boldsymbol{0}_{k}%
\end{pmatrix}%
\mathbb{E}\left[ \varepsilon \mid\boldsymbol{c}\right] =%
\begin{pmatrix}
z^{\ast }(\zeta _{0})\left( \varepsilon -\mathbb{E}\left[ \varepsilon \mid%
\boldsymbol{c}\right] \right)  \\ 
\boldsymbol{r}\varepsilon 
\end{pmatrix}%
$.}

The next result establishes consistency of the variance estimator.
\begin{theorem}\label{cons var IR} 
If the assumptions of Theorem \ref{Th asy distr} are satisfied, then
\[
\hat{\boldsymbol{G}}_{\beta}^{-1} \hat{\boldsymbol{\Omega}} (\hat{\boldsymbol{G}}_{\beta}')^{-1} \xrightarrow{p} \boldsymbol{G}_{\beta}^{-1} \boldsymbol{\Omega} (\boldsymbol{G}_{\beta}')^{-1}.
\]
\end{theorem}

\subsection{Asymptotics for the control function approach}

We now consider equation (\ref{ExM}) and define $\boldsymbol{q}^{\ast
}(\zeta )\coloneqq(z,\boldsymbol{r}^{\prime },\zeta (\boldsymbol{c}%
))^{\prime }$. Let $\boldsymbol{q}_{0}^{\ast }\coloneqq\boldsymbol{q}^{\ast
}(\zeta _{0})$ and, for $i=1,\ldots ,n$, $\boldsymbol{q}{_{i}^{\ast }}(\hat{%
\zeta})\coloneqq\bigl(z_{i},\boldsymbol{r}_{i}^{\prime },z_{i}^{\ast }(\hat{%
\zeta}\left( \boldsymbol{c}_{i}\right) )\bigr)^{\prime }$. The (two-step)
LIVE of $\boldsymbol{\beta }^{\ast }$ is 
\begin{equation*}
\boldsymbol{\hat{\beta}}^{\ast }\coloneqq\left( \sum_{i=1}^{n}\boldsymbol{q}{%
_{i}^{\ast }}(\hat{\zeta})\boldsymbol{x}_{i}^{\ast \prime }\right)
^{-1}\left( \sum_{i=1}^{n}\boldsymbol{q}{_{i}^{\ast }}(\hat{\zeta}%
)y_{i}\right).
\end{equation*}

The moment condition defining $\boldsymbol{\beta }_{0}^{\ast }$ is%
\begin{equation}
\mathbb{E}\left[ \boldsymbol{q}_{0}^{\ast }(y-\boldsymbol{x}^{\boldsymbol{%
\ast }\prime }\boldsymbol{\beta }_{0}^{\ast })\right] =\boldsymbol{0}_{k+2},
\label{mom cond CF}
\end{equation}%
and $\boldsymbol{\hat{\beta}}^{\ast }$ solves the sample moment condition 
\begin{equation}
\frac{1}{n}\sum_{i=1}^{n}\left( \boldsymbol{q}^{\ast }_i(\hat{\zeta})(y_i-%
\boldsymbol{x}^{\boldsymbol{\ast }\prime }_i\boldsymbol{\beta }^{\ast
})\right) =\boldsymbol{0}_{k+2}.  \label{sample mom cond CF}
\end{equation}

For the control function approach, we need to replace Assumption \ref{assumption Gbeta} with the following assumption.

\begin{assumptionstar}{\ref{assumption Gbeta}{\color{blue}$^{\ast}$}}
	\label{assumption G*beta}$\boldsymbol{G}_{\boldsymbol{\beta }}^{\ast }%
	\coloneqq\mathbb{E}[\boldsymbol{q}_{0}^{\ast }\boldsymbol{x}^{\ast \prime }]$
	is non singular.
\end{assumptionstar}

The control-function counterpart of Theorem 1 is as follows.

\begin{theorem}
\label{Th asy distr CF} If Assumptions \ref{assumption kernel}--\ref%
{assumption moments Y} and \ref{assumption G*beta} are satisfied, then 
\begin{equation*}
\sqrt{n}\left( \boldsymbol{\hat{\beta}}^{\ast }-\boldsymbol{\beta }%
_{0}^{\ast }\right) \overset{d}{\rightarrow }\mathcal{N}\left( 0,\boldsymbol{%
G}_{\boldsymbol{\beta }}^{\ast -1}\boldsymbol{\Omega }^{\ast }(\boldsymbol{G}%
_{\boldsymbol{\beta }}^{\ast \prime })^{-1}\right) ,
\end{equation*}%
where $\boldsymbol{\Omega ^{\ast }}\coloneqq\mathrm{var}\left( \boldsymbol{%
q_{0}^{\ast }}\varepsilon ^{\ast }+\boldsymbol{\varphi ^{\ast }}\right) $,
with 
\begin{equation*}
\boldsymbol{\varphi ^{\ast }}\coloneqq z^{\ast }({\zeta _{0}})%
\begin{pmatrix}
-{\phi_0 }\zeta _{0}(\boldsymbol{c}) \\ 
-\phi_0 \boldsymbol{r} \\ 
\mathbb{E}\left[ \varepsilon ^{\ast }\mid\boldsymbol{c}\right] -{\phi_0 }{\zeta
_{0}(\boldsymbol{c})}%
\end{pmatrix}%
.
\end{equation*}
\end{theorem}

The asymptotic variance $\boldsymbol{G}_{\boldsymbol{\beta ^{\ast }}}^{\ast
-1}\boldsymbol{\Omega ^{\ast }(}\boldsymbol{G}_{\boldsymbol{\beta ^{\ast }}%
}^{\ast \prime })^{-1}$ can be estimated by $\boldsymbol{\hat{G}}_{%
\boldsymbol{\beta ^{\ast }}}^{\ast -1}\boldsymbol{\hat{\Omega}^{\ast }(}%
\boldsymbol{\hat{G}}_{\boldsymbol{\beta ^{\ast }}}^{\ast \prime })^{-1}$,
where $\boldsymbol{\hat{G}_{\boldsymbol{\beta ^{\ast }}}^{\ast }}=\frac{1}{n}%
\sum_{i=1}^{n}\boldsymbol{q_{i}^{\ast }}(\hat{\zeta})\boldsymbol{\hat{x}}%
_{i}^{\ast \prime }$ and $\boldsymbol{\hat{\Omega}^{\ast }}=-\frac{1}{n}%
\sum_{i=1}^{n}\boldsymbol{\hat{\tau}^{\ast }}_{i}\boldsymbol{\hat{\tau}}%
_{i}^{\ast \prime }$, where $\boldsymbol{\hat{\tau}^{\ast }}_{i}\coloneqq%
\boldsymbol{q}{_{i}^{\ast }}(\hat{\zeta})(y_{i}-\boldsymbol{\hat{x}}%
_{i}^{\ast \prime }\boldsymbol{\hat{\beta}^{\ast }})+\boldsymbol{\hat{\varphi%
}_{i}^{\ast }}$, with 
\begin{equation*}
\boldsymbol{\hat{\varphi}_{i}^{\ast }}\coloneq z_{i}^{\ast }(\hat{\zeta}(\boldsymbol{%
c}_{i}))%
\begin{pmatrix}
-\hat{\phi}\hat{\zeta}(\boldsymbol{c}_{i}) \\ 
-\hat{\phi}\boldsymbol{r}_{i} \\ 
\hat{\varepsilon}_{i}^{\ast }-\hat{\phi}\hat{\zeta}(\boldsymbol{c}_{i})%
\end{pmatrix}%
,
\end{equation*}%
and $\hat{\varepsilon}_{i}^{\ast }\coloneqq y_{i}-\boldsymbol{\hat{x}}%
_{i}^{\ast \prime }\boldsymbol{\hat{\beta}}^{\ast }$, with $\hat{\boldsymbol{%
x}}_{i}^{\ast }\coloneqq(t_{i},\boldsymbol{r}_{i}^{\prime },\hat{\zeta}(%
\boldsymbol{c}_{i}))^{\prime }$.

\begin{theorem}\label{cons var CF} 
	If the assumptions of Theorem \ref{Th asy distr CF} are satisfied, then
	\[
	\boldsymbol{\hat{G}_{\beta}}^{*-1} \boldsymbol{\hat{\Omega}}^*(\boldsymbol{\hat{G}_{\beta}}^{*'})^{-1} \xrightarrow{p} \boldsymbol{G_{\beta}}^{*-1} \boldsymbol{\Omega}^{*} (\boldsymbol{G_{\beta}}^{*'})^{-1}.
	\]
\end{theorem}

It is interesting to notice that $\hat{\phi}$ plays an important role in the variance of the estimator. Since the value of $\hat{\phi}$ depends on the specification of $\boldsymbol{r}$, the choice of regressors to include in the equation to be estimated affects the asymptotic variance of the control-function estimator both through changing the variance of the errors and through its influence on the value of $\hat{\phi}$. This contrasts with the instrument-residual estimator where the choice of $\boldsymbol{r}$ only affects the variance of the estimator through the variance of the errors.

\section{\label{sec simul}Simulation evidence}

In this section we present the results of a small simulation study illustrating the behavior of the proposed approaches, focusing particularly on the effects of the curse of dimensionality. The key feature of the design we use is that the data generating process does not vary as we change the number of control variables used in the estimation. Specifically, in all cases considered, we use $d$ control variables, $c_{i1},\ldots ,c_{id}$, $i=1,\ldots ,n$, each of them drawn independently from a standard normal distribution. However, for all values of $d$ that we consider, these control variables enter the data generating process as $\tilde{c}_{i}\coloneqq \sum_{q=1}^{d}c_{iq}/\sqrt{d}$, which also follows a standard normal distribution. This allows us to generate data with the same distribution for any value of $d$, while allowing the number of control variables to vary. Having generated $c_{i1},\ldots ,c_{id}$ and $\tilde{c}_{i}$, for $i=1,\ldots ,n$, the remaining variables are generated as follows.

The instrument $z_{i}$ is generated as independent draws from a Bernoulli distribution with 
\begin{equation}
\Pr \left( z_{i}=1\mid c_{i1},\ldots ,c_{id}\right) =\Phi \left( \left( 1-\psi
\right) \tilde{c}_{i}+\psi \tilde{c}_{i}^{2}-\psi \right) ,  \label{zeta}
\end{equation}
where $\Phi \left( \cdot \right) $ denotes the normal cumulative distribution function. That is, for $\psi =0$, $\Pr \left(z_{i}=1\mid c_{i1},\ldots ,c_{id}\right) $ can be consistently estimated by a probit using $c_{i1},\ldots ,c_{id}$ as regressors \citep[as suggested by][]{lee2021,kim2024,lee2024}, but for other values of $\psi $ such model will be misspecified.

The treatment indicator is obtained as $t_{i}=\left( 1-z_{i}\right) t_{i}(0)+z_{i}t_{i}(1)$, where $t_{i}(0)$ and $t_{i}(1)$ denote potential treatment indicators and are constructed as
\begin{equation*}
t_{i}(0)=\mathds{1}\left[ \Phi \left( \frac{\tilde{c}_{i}+u_{i}}{\sqrt{2}}\right)
<\kappa _{\text{AT}}\right] ,\qquad t_{i}(1)=\mathds{1}\left[ \Phi \left( \frac{%
\tilde{c}_{i}+u_{i}}{\sqrt{2}}\right) <1-\kappa _{\text{NT}}\right] ,
\end{equation*}
where $\mathds{1}[a]$ is the usual indicator function of the event $a$, $u_{i}\sim \mathcal{N}\left( 0,1\right) $, and $0<\kappa _{\text{AT}},\kappa _{\text{NT}}<1$ denote the share of always takers and never takers, respectively, with $\kappa _{\text{AT}}+\kappa _{\text{NT}}\leq 1$. Note that $t_{i}(0)$ and $t_{i}(1)$ are both equal to $0$ with probability $\kappa _{\text{NT}}$ (the probability of being a never taker), are both $1$ with probability $\kappa _{\text{AT}}$ (the probability of being an always taker), and $t_{i}(1)>t_{i}(0)$ with probability $1-\kappa_{\text{AT}}-\kappa _{\text{NT}}$ (the probability of being a complier); there are no defiers.\footnote{Notice that the probability of being a complier determines the strength of the instrument because $t_{i}=z_{i}$ for compliers.}

Finally, the outcome $y_{i}$ is generated as
\begin{equation*}
y_{i}=\exp \left( 1 + \alpha _{i}t_{i} + \tilde{c}_{i} -0.2\tilde{c}_{i}^2 \right) +\eta _{i},
\end{equation*}%
where $\eta _{i}\sim \mathcal{N}\left( 0,1\right) $ and%
\begin{equation*}
\alpha _{i}=\alpha_{\text{NT}}\mathds{1}[ t_{i}(0)=t_{i}(1)=0] +\alpha_{%
\text{C}}\mathds{1}[ t_{i}(0)<t_{i}(1)] +\alpha_{\text{AT}}\mathds{1}[
t_{i}(0)=t_{i}(1)=1] ,
\end{equation*}
where $\alpha_{\text{NT}}$, $\alpha_{\text{C}}$, and $\alpha_{\text{AT}}$ denote the treatment effects for the subpopulations of never takers, compliers, and always takers, respectively. In this setup, endogeneity is caused by the fact that both $\alpha _{i}$ and $t_{i}$ depend on $u_{i}$. Additionally, there is also noise introduced through $\eta _{i}$ and through the misspecification of the equations being estimated.

Using this data generating process, we performed simulations for $d\in \left\{ 1,3,5,7,9\right\} $, $n\in \left\{ 500,2000,8000\right\} $, $\kappa _{\text{AT}}=\kappa _{\text{NT}}=0.25$, $\alpha _{\text{NT}}=-1$, $\alpha _{\text{C}}=0$, $\alpha _{\text{AT}}=1$, and for $\psi =1$, with new sets of variables being drawn independently for each Monte Carlo replication. In all cases, the goal is to estimate $\alpha _{\text{C}}$, which is the LATE, and the models are estimated using $c_{i1},\ldots ,c_{id}$ as controls.

We start by estimating the regression of $y_{i}$ on $t_{i}$ and $c_{i1},\ldots, c_{id}$ using $z_{i}$ as an instrument for $t_{i}$.\footnote{Therefore, in these simulations, $d=k$. We note that, as discussed before, the performance of the estimators may be improved by changing the functions of $c_{i1},\ldots, c_{id}$ used as explanatory variables in the equation, but we do not explore that possibility.} Additionally, we estimate $\alpha _{\text{C}}$ using \citeauthor{lee2024}'s (\citeyear{lee2024}) \texttt{psr} Stata command (with the default options), which implements the estimator proposed by \cite{lee2021},\footnote{This is essentially the LIVE of $y_{i}$ on $t_{i}$ only, using a residual instrument based on an estimate of $\mathbb{E}\left[ z_{i}\mid c_{i1},\ldots, c_{id}\right] $ obtained with a probit. Notice that, because we set $\psi =1$, this estimator is based on a misspecified model for $\mathbb{E}\left[ z_{i}\mid c_{i1},\ldots, c_{id}\right] $.} as well as using \citeauthor{ahrens2024}'s (\citeyear{ahrens2024}) \texttt{ddml} Stata command, which implements the DML estimator of the partially linear instrumental variables model presented by \cite{chernozhukov2018}. In light of the discussion in Section \ref{sec approaches}, the nonparametric estimations needed to implement the DML are performed using the same neural network estimator we use in the implementation of the methods we propose; see more details below. We also run some simulations in which the nonparametric estimates needed for the DML estimator were obtained using an ensemble method of the type used by \cite{blandhol2025}, but the performance of the estimator was substantially worse in that case.

For the implementation of the proposed estimators, we considered their oracle version in which the preliminary step uses $\zeta _{0}(\boldsymbol{c}) $, which is given by (\ref{zeta}), and two feasible versions based on nonparametric estimates of $\zeta _{0}(\boldsymbol{c})$. Specifically, we estimated $\zeta _{0}(\boldsymbol{c})$ both by using the standard Nadaraya-Watson kernel estimator and a neural network regression. In line with the results in Section \ref{sec asy}, the kernel estimator is implemented using multiplicative Epanechnikov kernels of order $d+1$.\footnote{See \cite{Hansen2005} for details on how to construct such kernels.} After some experimentation, the bandwidth for each regressor was set to $(1.1+0.725d)\hat\sigma n^{\frac{-1}{2d+1}}$, where $\hat\sigma$ denotes the estimated standard deviation of the regressor.\footnote{We let the scaling constant increase with the dimension of $\boldsymbol{c}$, and therefore with the order of the kernel, to avoid overfitting when $d$ increases. It may be possible to improve the performance of the estimator by using a different scaling constant, but we did not explore that further. In practice, the scaling constant needs to be chosen on a case-by-case basis.} As mentioned before, the neural network estimator we use is the one implemented by \cite{ahrens2023} in their \texttt{pystacked} Stata command, with the default options, which is based on the ReLU activation function, has a single hidden layer with $100$ nodes, and uses the Adam optimizer proposed by \cite{kingma2014} for weight optimization.\footnote{We also implemented the proposed estimators  based on the neural network regressions using the cross-fitting approach suggested by \cite{chernozhukov2018}. However, that did not improve the performance of the estimators \citep[see][]{farrell2021}.}

Table 1 summarizes the results for each case and for each estimator. Specifically, the table reports the median of the estimates obtained in $1000 $ replications of the simulation procedure, as well as the interquartile range (in parentheses).\footnote{We report the median and the interquartile range because it is well known that the finite sample distribution of the exactly-identified LIVE has no finite moments, see \cite{kinal1980}.} Because $\alpha _{\text{C}}=0$, these results can also be interpreted as the median and interquartile range of the bias of the estimators. In the table, the results of the instrument-residual and control-function estimators are labeled Oracle when they are obtained using the true conditional expectations, and labeled NW and NN when the conditional expectations are obtained, respectively, with the Nadaraya-Watson and neural network estimators. 

The results in Table 1 show that, as expected, the LIVE does not identify the parameter of interest. The results obtained with \citeauthor{lee2021}'s (\citeyear{lee2021}) estimator are more interesting. This estimator is based on a misspecified model for $\zeta _{0}(\boldsymbol{c})$, but the method implemented by \cite{lee2024} offers some protection against misspecification of this function and, for $d=1$, this approach has substantially smaller bias than the LIVE. However, as $d$ increases, the effects of the misspecification become evident, and the results obtained with this estimator become similar to those obtained with the LIVE. 

The DML estimator performs very well for the larger sample sizes used in these experiments. Indeed, for the five values of $d$ that we consider, DML has essentially no bias when $n\in \left\{2000,8000\right\}$. However, for $n = 500$, DML has noticeable biases, especially for larger values of $d$. The performance of the DML estimator is better in these simulations than in the ones reported by \cite{blandhol2025}, who note that the DML estimator tends to have a sizable bias that vanishes slowly as the sample size increases. This difference may be a consequence of the different simulation designs, but may also result from the different ways in which the estimator is implemented; as noted before, we also performed experiments in which the nonparametric estimates were obtained using an ensemble method as in \cite{blandhol2025}, but the performance of the estimator was worse in that case.

Turning now to the results obtained with the estimators proposed in this paper, we start by noting that the performance of the oracle estimators does not depend on $d$, which is to be expected because these estimators are not affected by the curse of dimensionality. Additionally, the two oracle estimators have nearly identical performances, presenting essentially no bias and similar levels of dispersion, which drops with the square root of the sample size. \renewcommand{\arraystretch}{.95}

As for the feasible estimators, we find that results obtained with the instrument-residual estimator are generally better than the ones based on the control-function approach. This is especially noticeable for $d$ larger than $1$ because the performance of control-function estimator is much more sensitive to the value of $d$ than that of the instrument-residual estimator. Moreover, we find that the estimators that use the neural network estimate of $\zeta _{0}(\boldsymbol{c})$ (whose results are labeled NN) tend to outperform the estimators based on the Nadaraya-Watson kernel regression (whose results are labeled NW), with the difference being particularly clear in the case of the control-function approach. Finally, we note that, as for the oracle estimators, the dispersion of the estimates obtained with the feasible estimators drops with the square root of the sample size.

Overall, these results suggest that, for small $d$, there is little to choose between the four versions of the estimators we introduced. However, for larger values of $d$, the estimators based on the neural network estimates of the conditional expectation of $z_{i}$ appear to dominate their competitors, with the instrument-residual estimator performing particularly well. Indeed, the results obtained with the instrument-residual estimator that uses a neural network estimator in the preliminary step are remarkably insensitive to the value of $d$ and the performance of this estimator, both in terms of bias and dispersion, is comparable to that of the oracle estimators. This estimator is closely related to the DML estimator and has a similar performance in the larger samples, but generally outperforms it for the smaller samples.


\begin{center}
$%
\begin{tabular}{ccccccccccccc}
\multicolumn{13}{c}{Table 1: Simulation results} \\ \hline
&  &  &  &  &  & \multicolumn{3}{c}{Control Function} &  & 
\multicolumn{3}{c}{Instrument Residual} \\ \cline{7-9}\cline{11-13}
$d$ &  & LIVE & PSR & DML &  & Oracle & NW & NN &  & Oracle & NW & NN \\ 
\hline
\multicolumn{13}{c}{$n=500$} \\ \hline
$1$ &  & \multicolumn{1}{r}{$\underset{\left( 0.43\right) }{0.90}$} & 
\multicolumn{1}{r}{$\underset{\left( 0.40\right) }{0.18}$} & 
\multicolumn{1}{r}{$\underset{\left( 0.53\right) }{0.01}$} &  & 
\multicolumn{1}{r}{$-\underset{\left( 0.52\right) }{0.00}$} & 
\multicolumn{1}{r}{$-\underset{\left( 0.53\right) }{0.01}$} & 
\multicolumn{1}{r}{$\underset{\left( 0.51\right) }{0.02}$} &  & 
\multicolumn{1}{r}{$\underset{\left( 0.54\right) }{0.01}$} & 
\multicolumn{1}{r}{$\underset{\left( 0.53\right) }{0.03}$} & 
\multicolumn{1}{r}{$\underset{\left( 0.51\right) }{0.05}$} \\ 
$3$ &  & \multicolumn{1}{r}{$\underset{\left( 0.40\right) }{0.86}$} & 
\multicolumn{1}{r}{$\underset{\left( 0.48\right) }{0.74}$} & 
\multicolumn{1}{r}{$-\underset{\left( 0.51\right) }{0.07}$} &  & 
\multicolumn{1}{r}{$-\underset{\left( 0.50\right) }{0.02}$} & 
\multicolumn{1}{r}{$\underset{\left( 0.59\right) }{0.11}$} & 
\multicolumn{1}{r}{$-\underset{\left( 0.52\right) }{0.05}$} &  & 
\multicolumn{1}{r}{$-\underset{\left( 0.49\right) }{0.01}$} & 
\multicolumn{1}{r}{$-\underset{\left( 0.56\right) }{0.03}$} & 
\multicolumn{1}{r}{$-\underset{\left( 0.51\right) }{0.00}$} \\ 
$5$ &  & \multicolumn{1}{r}{$\underset{\left( 0.40\right) }{0.87}$} & 
\multicolumn{1}{r}{$\underset{\left( 0.43\right) }{0.82}$} & 
\multicolumn{1}{r}{$-\underset{\left( 0.50\right) }{0.10}$} &  & 
\multicolumn{1}{r}{$\underset{\left( 0.48\right) }{0.02}$} & 
\multicolumn{1}{r}{$\underset{\left( 0.66\right) }{0.32}$} & 
\multicolumn{1}{r}{$-\underset{\left( 0.52\right) }{0.04}$} &  & 
\multicolumn{1}{r}{$\underset{\left( 0.48\right) }{0.02}$} & 
\multicolumn{1}{r}{$-\underset{\left( 0.82\right) }{0.07}$} & 
\multicolumn{1}{r}{$\underset{\left( 0.50\right) }{0.05}$} \\ 
$7$ &  & \multicolumn{1}{r}{$\underset{\left( 0.44\right) }{0.88}$} & 
\multicolumn{1}{r}{$\underset{\left( 0.46\right) }{0.83}$} & 
\multicolumn{1}{r}{$-\underset{\left( 0.53\right) }{0.15}$} &  & 
\multicolumn{1}{r}{$-\underset{\left( 0.50\right) }{0.01}$} & 
\multicolumn{1}{r}{$\underset{\left( 0.77\right) }{0.07}$} & 
\multicolumn{1}{r}{$-\underset{\left( 0.56\right) }{0.13}$} &  & 
\multicolumn{1}{r}{$-\underset{\left( 0.51\right) }{0.01}$} & 
\multicolumn{1}{r}{$\underset{\left( 0.81\right) }{0.02}$} & 
\multicolumn{1}{r}{$\underset{\left( 0.52\right) }{0.03}$} \\ 
$9$ &  & \multicolumn{1}{r}{$\underset{\left( 0.42\right) }{0.89}$} & 
\multicolumn{1}{r}{$\underset{\left( 0.43\right) }{0.86}$} & 
\multicolumn{1}{r}{$-\underset{\left( 0.53\right) }{0.17}$} &  & 
\multicolumn{1}{r}{$\underset{\left( 0.49\right) }{0.00}$} & 
\multicolumn{1}{r}{$-\underset{\left( 1.21\right) }{0.19}$} & 
\multicolumn{1}{r}{$-\underset{\left( 0.57\right) }{0.18}$} &  & 
\multicolumn{1}{r}{$-\underset{\left( 0.51\right) }{0.00}$} & 
\multicolumn{1}{r}{$\underset{\left( 0.89\right) }{0.09}$} & 
\multicolumn{1}{r}{$\underset{\left( 0.51\right) }{0.05}$} \\ \hline
\multicolumn{13}{c}{$n=2000$} \\ \hline
$1$ &  & \multicolumn{1}{r}{$\underset{\left( 0.20\right) }{0.89}$} & 
\multicolumn{1}{r}{$\underset{\left( 0.19\right) }{0.18}$} & 
\multicolumn{1}{r}{$\underset{\left( 0.25\right) }{0.01}$} &  & 
\multicolumn{1}{r}{$\underset{\left( 0.25\right) }{0.01}$} & 
\multicolumn{1}{r}{$-\underset{\left( 0.25\right) }{0.00}$} & 
\multicolumn{1}{r}{$\underset{\left( 0.26\right) }{0.01}$} &  & 
\multicolumn{1}{r}{$\underset{\left( 0.25\right) }{0.01}$} & 
\multicolumn{1}{r}{$\underset{\left( 0.25\right) }{0.01}$} & 
\multicolumn{1}{r}{$\underset{\left( 0.26\right) }{0.02}$} \\ 
$3$ &  & \multicolumn{1}{r}{$\underset{\left( 0.22\right) }{0.89}$} & 
\multicolumn{1}{r}{$\underset{\left( 0.34\right) }{0.77}$} & 
\multicolumn{1}{r}{$-\underset{\left( 0.27\right) }{0.00}$} &  & 
\multicolumn{1}{r}{$\underset{\left( 0.27\right) }{0.00}$} & 
\multicolumn{1}{r}{$\underset{\left( 0.34\right) }{0.10}$} & 
\multicolumn{1}{r}{$-\underset{\left( 0.28\right) }{0.00}$} &  & 
\multicolumn{1}{r}{$\underset{\left( 0.27\right) }{0.00}$} & 
\multicolumn{1}{r}{$-\underset{\left( 0.28\right) }{0.01}$} & 
\multicolumn{1}{r}{$\underset{\left( 0.27\right) }{0.00}$} \\ 
$5$ &  & \multicolumn{1}{r}{$\underset{\left( 0.22\right) }{0.89}$} & 
\multicolumn{1}{r}{$\underset{\left( 0.25\right) }{0.83}$} & 
\multicolumn{1}{r}{$-\underset{\left( 0.26\right) }{0.01}$} &  & 
\multicolumn{1}{r}{$-\underset{\left( 0.26\right) }{0.01}$} & 
\multicolumn{1}{r}{$\underset{\left( 0.38\right) }{0.62}$} & 
\multicolumn{1}{r}{$-\underset{\left( 0.27\right) }{0.05}$} &  & 
\multicolumn{1}{r}{$-\underset{\left( 0.26\right) }{0.01}$} & 
\multicolumn{1}{r}{$-\underset{\left( 0.51\right) }{0.06}$} & 
\multicolumn{1}{r}{$-\underset{\left( 0.26\right) }{0.02}$} \\ 
$7$ &  & \multicolumn{1}{r}{$\underset{\left( 0.23\right) }{0.88}$} & 
\multicolumn{1}{r}{$\underset{\left( 0.23\right) }{0.86}$} & 
\multicolumn{1}{r}{$-\underset{\left( 0.26\right) }{0.01}$} &  & 
\multicolumn{1}{r}{$-\underset{\left( 0.25\right) }{0.01}$} & 
\multicolumn{1}{r}{$\underset{\left( 0.46\right) }{0.36}$} & 
\multicolumn{1}{r}{$-\underset{\left( 0.26\right) }{0.07}$} &  & 
\multicolumn{1}{r}{$-\underset{\left( 0.25\right) }{0.00}$} & 
\multicolumn{1}{r}{$-\underset{\left( 0.50\right) }{0.06}$} & 
\multicolumn{1}{r}{$-\underset{\left( 0.25\right) }{0.01}$} \\ 
$9$ &  & \multicolumn{1}{r}{$\underset{\left( 0.20\right) }{0.88}$} & 
\multicolumn{1}{r}{$\underset{\left( 0.22\right) }{0.86}$} & 
\multicolumn{1}{r}{$-\underset{\left( 0.26\right) }{0.00}$} &  & 
\multicolumn{1}{r}{$-\underset{\left( 0.25\right) }{0.01}$} & 
\multicolumn{1}{r}{$-\underset{\left( 0.44\right) }{0.07}$} & 
\multicolumn{1}{r}{$-\underset{\left( 0.26\right) }{0.12}$} &  & 
\multicolumn{1}{r}{$-\underset{\left( 0.26\right) }{0.00}$} & 
\multicolumn{1}{r}{$-\underset{\left( 0.41\right) }{0.00}$} & 
\multicolumn{1}{r}{$-\underset{\left( 0.26\right) }{0.02}$} \\ \hline
\multicolumn{13}{c}{$n=8000$} \\ \hline
$1$ &  & \multicolumn{1}{r}{$\underset{\left( 0.11\right) }{0.88}$} & 
\multicolumn{1}{r}{$\underset{\left( 0.11\right) }{0.19}$} & 
\multicolumn{1}{r}{$-\underset{\left( 0.13\right) }{0.01}$} &  & 
\multicolumn{1}{r}{$-\underset{\left( 0.13\right) }{0.01}$} & 
\multicolumn{1}{r}{$-\underset{\left( 0.13\right) }{0.01}$} & 
\multicolumn{1}{r}{$\underset{\left( 0.13\right) }{0.00}$} &  & 
\multicolumn{1}{r}{$-\underset{\left( 0.13\right) }{0.01}$} & 
\multicolumn{1}{r}{$-\underset{\left( 0.13\right) }{0.00}$} & 
\multicolumn{1}{r}{$-\underset{\left( 0.13\right) }{0.00}$} \\ 
$3$ &  & \multicolumn{1}{r}{$\underset{\left( 0.11\right) }{0.88}$} & 
\multicolumn{1}{r}{$\underset{\left( 0.24\right) }{0.79}$} & 
\multicolumn{1}{r}{$\underset{\left( 0.13\right) }{0.00}$} &  & 
\multicolumn{1}{r}{$-\underset{\left( 0.13\right) }{0.00}$} & 
\multicolumn{1}{r}{$\underset{\left( 0.20\right) }{0.06}$} & 
\multicolumn{1}{r}{$-\underset{\left( 0.13\right) }{0.00}$} &  & 
\multicolumn{1}{r}{$-\underset{\left( 0.13\right) }{0.00}$} & 
\multicolumn{1}{r}{$-\underset{\left( 0.14\right) }{0.00}$} & 
\multicolumn{1}{r}{$-\underset{\left( 0.13\right) }{0.00}$} \\ 
$5$ &  & \multicolumn{1}{r}{$\underset{\left( 0.10\right) }{0.89}$} & 
\multicolumn{1}{r}{$\underset{\left( 0.14\right) }{0.85}$} & 
\multicolumn{1}{r}{$\underset{\left( 0.14\right) }{0.01}$} &  & 
\multicolumn{1}{r}{$\underset{\left( 0.14\right) }{0.01}$} & 
\multicolumn{1}{r}{$\underset{\left( 0.22\right) }{0.77}$} & 
\multicolumn{1}{r}{$-\underset{\left( 0.14\right) }{0.00}$} &  & 
\multicolumn{1}{r}{$\underset{\left( 0.14\right) }{0.01}$} & 
\multicolumn{1}{r}{$-\underset{\left( 0.30\right) }{0.04}$} & 
\multicolumn{1}{r}{$\underset{\left( 0.14\right) }{0.00}$} \\ 
$7$ &  & \multicolumn{1}{r}{$\underset{\left( 0.10\right) }{0.88}$} & 
\multicolumn{1}{r}{$\underset{\left( 0.13\right) }{0.86}$} & 
\multicolumn{1}{r}{$-\underset{\left( 0.13\right) }{0.02}$} &  & 
\multicolumn{1}{r}{$-\underset{\left( 0.13\right) }{0.00}$} & 
\multicolumn{1}{r}{$\underset{\left( 0.20\right) }{0.78}$} & 
\multicolumn{1}{r}{$-\underset{\left( 0.13\right) }{0.04}$} &  & 
\multicolumn{1}{r}{$-\underset{\left( 0.13\right) }{0.00}$} & 
\multicolumn{1}{r}{$-\underset{\left( 0.42\right) }{0.13}$} & 
\multicolumn{1}{r}{$-\underset{\left( 0.13\right) }{0.01}$} \\ 
$9$ &  & \multicolumn{1}{r}{$\underset{\left( 0.10\right) }{0.88}$} & 
\multicolumn{1}{r}{$\underset{\left( 0.12\right) }{0.86}$} & 
\multicolumn{1}{r}{$-\underset{\left( 0.14\right) }{0.01}$} &  & 
\multicolumn{1}{r}{$-\underset{\left( 0.14\right) }{0.01}$} & 
\multicolumn{1}{r}{$\underset{\left( 0.40\right) }{0.28}$} & 
\multicolumn{1}{r}{$-\underset{\left( 0.14\right) }{0.07}$} &  & 
\multicolumn{1}{r}{$-\underset{\left( 0.14\right) }{0.01}$} & 
\multicolumn{1}{r}{$-\underset{\left( 0.25\right) }{0.06}$} & 
\multicolumn{1}{r}{$-\underset{\left( 0.14\right) }{0.02}$} \\ \hline
\multicolumn{13}{p{14.8cm}}{{\footnotesize \textbf{Note:} The table reports
the median and interquartile range (in parentheses) of the bias for the
LIVE, PSR, and DML estimators, as well as for the Instrument-Residual and
Control-Function estimators obtained using different estimates of the
conditional expectation of $z$. The results labeled Oracle are obtained with
the true conditional expectation, whereas for those labeled NW and NN the
conditional expectation is obtained, respectively, with the kernel and
neural network estimators.}}%
\end{tabular}%
\bigskip $
\end{center}
\newpage

The advantage of the neural network-based estimators may be an artifact of the way the different estimators are implemented or of the simulation design used. However, as discussed before, there are reasons to believe that neural network estimators are able to circumvent the curse of dimensionality \citep[see,][]{bach2017,bauer2019,schmidt2020,kohler2021,braun2024}, which may explain their good performance in this exercise. As for the choice between the two approaches that we introduced, the instrument-residual estimator appears to be a safer bet, both because it appears to be less sensitive to the choice to the preliminary nonparametric estimator, but also because it generally has a smaller bias than the control-function estimator.

\section{\label{sec examples}An illustrative application}

In this section we illustrate the application of the proposed estimators by
revisiting the work of \cite{dube2020queens}, an empirical application
highlighted by \cite{blandhol2025} in which both the treatment and
instrument are binary.

Using data from 1480 to 1913, \cite{dube2020queens} investigated whether
European states experienced more peace under female leadership. In their
study, the outcome of interest is a binary indicator for whether a polity
was at war in a given year and the treatment variable is a dummy for whether
the polity was ruled by a queen in that year. 

\cite{dube2020queens} consider both just- and over-identified estimators, and here we focus on the just-identified case where the instrument is a binary indicator of whether the previous monarch had a legitimate firstborn male child.\footnote{%
The results obtained by \cite{dube2020queens} for the just- and
over-identified cases are broadly similar.} To address concerns about the
instrument's validity, \cite{dube2020queens} control for a set of covariates
that includes polity and decade fixed effects, an indicator for whether the
previous monarchs were unrelated co-rulers, a binary variable for whether the gender of the firstborn of previous monarchs is missing, an indicator for whether previous monarchs had at least one legitimate child and the birth year is known, and a similar indicator for when the birth year is unknown. In total, there are $64$ control variables, all of which are dummies. Both \cite{dube2020queens} and \cite{blandhol2025} emphasize the importance of controlling for the covariates in this analysis.

Column (1) of Table 2 reproduces the estimated effects (and cluster-robust standard errors) reported by \cite{blandhol2025}, which differ only slightly from those in the original paper. The difference between the LIVE results with and without covariates confirms the importance of accounting for the role of the controls, while the difference between the least squares and LIVE results suggests the need to account for the endogeneity of the treatment. A remarkable feature of these results is that the DML estimate is qualitatively similar to that obtained with the LIVE with covariates. This is surprising because the results of \citeauthor{ramsey1969tests}'s (\citeyear{ramsey1969tests}) RESET test reported by \cite{blandhol2025} suggest that the LIVE with covariates does not fulfill the rich-covariates condition, and therefore we would expect it to deliver an estimate reasonably different from the one obtained with DML,  because DML is supposed to identify $\alpha_{\text{rich}}$ but the LIVE is not. We will return to this point soon.

\bigskip 

\begin{center}
\begin{tabular}{lcc}
\multicolumn{3}{c}{Table 2 : Estimation results} \\ \hline\hline
& (1) & (2) \\ \hline
Least squares & \multicolumn{1}{r}{$0.115$ $(0.035)$} & \multicolumn{1}{r}{$0.115$ $%
(0.035)$} \\ 
LIVE, no covariates & \multicolumn{1}{r}{$1.011$ $(0.522)$} & 
\multicolumn{1}{r}{$1.011$ $(0.522)$} \\ 
LIVE, with covariates & \multicolumn{1}{r}{$0.400$ $(0.211)$} & 
\multicolumn{1}{r}{$0.400$ $(0.211)$} \\ 
DML ensemble & \multicolumn{1}{r}{$0.318$ $(0.240)$} & 
\multicolumn{1}{r}{$0.382$ $(0.395)$} \\ 
DML neural net no cross-fitting & \multicolumn{1}{r}{} & \multicolumn{1}{r}{$%
-0.518$ $(0.498)$} \\ 
IR neural net no covariates& \multicolumn{1}{r}{} & \multicolumn{1}{r}{$-0.490$ $%
(1.088)$} \\ 
IR neural net & \multicolumn{1}{r}{} & \multicolumn{1}{r}{$-0.592$ $(0.641)$}\\ 
CF neural net & \multicolumn{1}{r}{} & \multicolumn{1}{r}{$-0.438$ $(0.527)$}
\\  \hline\hline
\multicolumn{3}{p{11.8cm}}{{\footnotesize \textbf{Note:} The table reports
estimates of the effect of queen rule on the probability of war obtained by different methods. The sample size is $3586$ and there are $64$ covariates.}}
\end{tabular}
\end{center}
\bigskip 
\newpage
Column (2) of Table 2 presents our own estimates, which match those in column
(1) for the least squares and the LIVE. For the DML, our estimate is not identical to the one reported by \cite{blandhol2025}, but the difference is small enough to be entirely justified by the fact that the estimator is based on random sample splitting.\footnote{Like \cite{blandhol2025}, we report the median of the estimates over $100$ repetitions, form folds based on the clustering indicator, and perform the preliminary step using the same ensemble estimator, but the implementation of the estimator may differ in other aspects not explicitly mentioned by \cite{blandhol2025}.}

The four bottom rows of Column (2) in Table 2 report results obtained with different estimators in which the relevant non-parametric regressions are performed using the same neural network estimator that we used in the simulations. The estimators used to obtain these results are, respectively, a DML-type estimator that does not use cross-fitting,\footnote{Alternatively, this estimator can be seen as an instrumental residual estimator in which $\boldsymbol{r}$ is composed of estimates of $\mathbb{E}\left[ y\mid \boldsymbol{c}\right] $ and $\mathbb{E}\left[ t\mid \boldsymbol{c}\right] $, also obtained using the shallow neural network estimator.} the instrument-residual estimator proposed earlier but without including covariates in the equation, as in \cite{lee2021}, and the proposed instrument-residual and control-function estimators that include all covariates in the equation. None of these estimators uses cross-fitting and therefore each regression was estimated only once. The reported standard errors should be seen only as indicative because they do not account for the fact that the estimator relies on a preliminary first step.\footnote{Specifically, we report standard instrumental variables  clustered standard errors, using the same $176$ clusters considered by \cite{dube2020queens}.} 

The results obtained with these estimators are strikingly different from the
ones obtained before because they have the opposite sign. Except for least squares, the standard errors associated with all these estimates are relatively large, and therefore one may think that the different signs are just the result of sampling noise. However, the fact that there are no sign reversals in each half of the table suggests that there may a deeper reason for this.

Because all covariates in this application are binary, it is possible to obtain information on $\alpha_{\text{rich}}$ by estimating a saturated model, and \cite{blandhol2022} report that the estimate obtained with the saturated model is $-0.509$ $(0.523)$. This suggests that, with rich-covariates, we should expect negative estimates of the effect, such as those obtained with the methods we proposed and reported in the bottom part of column (2) of Table 2. Therefore, what is puzzling is that the standard DML estimator, whose results are reported in the fourth line of Table 2, leads to estimates that are closer to the one obtained with the LIVE without rich covariates, than to the one obtained with the saturated model. The explanation for this turns out to be quite simple.

The cross-fitting used in DML accounts for the panel nature of the data by forming sample splits based on the variable used to cluster the standard errors. This ensures that the folds are independent. However, because many of the dummies used as controls are rarely equal to $1$, it may not be possible to identify their effect in some of the sub-samples.\footnote{For example, the mean of the dummy indicator of whether the previous monarchs were unrelated co-rulers is $0.008$ and this variable is only equal to $1$ in three clusters. Therefore, if these clusters are not included in a particular sub-sample, it is not possible to identify the effect of this variable.} This implies that the conditional expectations estimated by cross-fitting only control for covariates with variation in the relevant sub-sample. Consequently, these are not valid estimates of the required conditional expectations, and therefore such estimators do not identify $\alpha _{\text{rich}}$. This problem seriously restricts the practical usefulness of estimators based on cross-fitting in applications involving sparse binary covariates.

Another telltale sign of this problem is that the DML estimates based on cross fitting are very sensitive to the sample split, reflecting the fact that the set of variables effectively controlled for varies with the way the sample is split.\footnote{In the $100$ sample splits we used to obtain the value reported in Table 2, the estimates vary between $ -0.202$ and $2.421$.} To mitigate this, and as in \cite{blandhol2025}, the DML result we report is the median of the estimates obtained in $100$ repetitions of the estimation process. This increases the computational cost of the estimator very substantially: on a standard desktop computer, estimation with the proposed instrument-residual and control-function methods took less than $30$ seconds, while the DML estimates took over $18$ hours.  The immense increase in computational cost, however,  does not eliminate the fundamental flaw of this method,  which still delivers an unreliable estimate.
 
 The point estimates obtained with the four methods that do not use cross-fitting are relatively close to each other, and are also close the estimate obtained with the saturated model, suggesting that in this example all these methods provide reasonable estimates of $\alpha _{\text{rich}}$. Although, as noted before, the reported standard errors do not account for the variability introduced by the first-step estimation, these results point to potential significant efficiency  gains from including the covariates in the equation to be estimated, especially if the estimator controls for the covariates in a flexible way.

In summary, in this application, the proposed instrument-residual and control-function estimators deliver results close to the one obtained with a saturated model, which suggests that both methods are successful in ensuring that the rich-covariates condition is satisfied. Therefore, when it is impossible or impractical to estimate saturated models, the methods we propose may be an appealing alternative; these estimators also do not suffer from the drawback of the DML estimators that this application highlighted. Moreover, the results obtained with the proposed estimators suggest that, in this application, violation of the rich-covariates condition leads to an estimate with the ``wrong sign,'' a phenomenon emphasized by \cite{blandhol2025}.\footnote{See their numerical illustration, presented in Subsection 2.3.} Therefore, in contrast with the findings of \cite{dube2020queens}, our results suggest that states led by queens engaged in war less than those led by kings, although this effect is imprecisely estimated. Using over-identified estimators may help to more accurately identify the effect of queens on war, but we do not pursue that avenue here.

\section{\label{sec concl}Concluding remarks}

For many years, linear instrumental variables estimators have been one of
the main tools used by applied economists to estimate causal relationships.
Specifically, guided by the findings of \cite{Imbens1994} and
\cite{angrist1995}, practitioners often use instrumental variables
estimators with the aim of identifying the average treatment effect on the
population of compliers, the so-called LATE. However, the ability of such
estimators to identify causal relationships has gradually been called into
question, and there is a widespread view that, in models with covariates,
linear instrumental variables estimators only identify a causal effect if
the model is saturated, in the sense that it includes a dummy variable for
each possible combination of the values of the covariates \citep[see,][]{mogstad2024, blandhol2025}.

Because saturated models are often impractical or even infeasible, it is interesting to consider alternative approaches to the estimation of causal effects in situations where it is important to account for the role of covariates. \cite{lee2021} and \cite{kim2024} introduced linear instrumental variables estimators that do not rely on saturation and can identify causal relationships in models with covariates. These estimators depend on a preliminary step that ensures that the resulting estimand has a causal interpretation. However, these estimators are based on a strong parametric assumption that may often be invalid, and therefore are not particularly
attractive. 

\cite{kim2024} and \cite{blandhol2025} point out that one way to estimate causal effects without the need to rely on parametric assumptions is to use the double/debiased machine learning estimator of the partially linear instrumental variables model considered by \cite{chernozhukov2018}; see also \cite{LeeandLee25}. However, the empirical illustration presented in Section \ref{sec examples} revealed that this estimator may not be able to ensure the fulfillment of rich-covariates condition when the controls include regressors with little variation, such as dummy variables that are rarely equal to $1$. This is a serious problem that restricts the applicability of the double/debiased machine learning estimator.

We propose two versions of the linear instrumental variables estimator, the
instrument-residual and control-function estimators, that retain most of the
simplicity of the methods proposed by \cite{lee2021} and \cite{kim2024} but,
like the estimators of \cite{chernozhukov2018}, do not rely on parametric
assumptions. \cite{BH23} propose similar estimators to address a different but related problem. Importantly, the estimators we propose have an estimand with a
clear causal interpretation, are easy to implement, and are closely related
to estimators practitioners are familiar with. 

The two estimators that we propose depend on a preliminary step, in which
the expectation of the instrument conditional on the covariates is estimated
nonparametrically. We show that the estimators of the parameter of interest
can converge at the usual parametric rate, even when the estimator in the
preliminary step converges at a slower rate. We prove this result for the
case where the preliminary step is performed using a kernel regression, but
it may be possible to obtain similar results when other nonparametric
methods are used in this step.

Traditional methods to estimate conditional expectations, such as kernel and
series regressions, can be severely affected by the curse of dimensionality,
and therefore may be unsuitable in many practical situations. Indeed, the
simulation evidence we present in Section \ref{sec simul} shows that, when a
kernel regression is used in the preliminary step, the performance of the
estimators deteriorates quickly with the number of controls, with the
control-function estimator being particularly affected by the curse of
dimensionality.

The current research on machine learning methods may provide attractive alternatives to the use of traditional nonparametric methods and, as discussed in Section \ref{sec approaches}, there is evidence that neural network estimators can offer some robustness to the curse of dimensionality. This is borne out by our simulation results in which the estimators that use a neural network in the first step are remarkably insensitive to the number of covariates in the regression---especially the instrument-residual estimator. The empirical illustrations in Section \ref{sec examples} also suggest that the method works well in practice.

Given how active the research on machine learning methods is, it would not be prudent to advocate the use of a particular estimation method to perform the preliminary step. However, based on our reading of the literature and on the simulation results we report, we recommend that, at least for now, practitioners perform the estimation of the preliminary step using a shallow neural network based on the ReLU activation function, with a large number of nodes, and using the Adam optimizer for weight optimization.

There are a number of extensions of the methods we propose that would be interesting to explore. Our simulation results suggest that the proposed estimators perform particularly well when the preliminary step uses a neural network, but we only provide results on the asymptotic distribution of the estimators whose preliminary step is performed using kernel regression. Therefore, establishing asymptotic properties for estimators that use other nonparametric methods is a priority. Moreover, we only explicitly considered the case of just identified models with a binary treatment and a binary instrument. Using the results in \cite{kim2024} and \cite{BH23}, our results can easily be extended to other types of instrument and to overidentified models, but it may also be possible to formally extend them to the case where there are multiple treatments. Finally, it would be also very interesting to have more precise guidance on how to perform the nonparametric estimation in the preliminary step, and on the role that machine learning methods can have on reducing the impact of the curse of dimensionality.

\begin{appendices}

\section{\label{App estimands}The estimands} If Assumption \ref{assumption
Gbeta} is satisfied, the unique solution to moment condition (\ref{mom cond
IR}) is 
\begin{equation*} \underset{(k+1)\times 1}{\boldsymbol{\beta }_{0}}%
=(\alpha _{0},\boldsymbol{ \gamma }_{0}^{\prime })^{\prime }=\left( \mathbb{%
E}[\boldsymbol{q}_{0} \boldsymbol{x}^{\prime }]\right) ^{-1}\mathbb{E[}%
\boldsymbol{q}_{0}y]. \end{equation*}
Similarly, if Assumption \ref{%
assumption G*beta} is satisfied, the unique solution to moment condition (%
\ref{mom cond CF}) is
\begin{equation*} \underset{(k+2)\times 1}{\boldsymbol{%
\beta }_{0}^{\ast }}=(\alpha _{0}^{\ast },\boldsymbol{\gamma }_{0}^{\ast
\prime },\phi _{0})^{\prime }=\left( \mathbb{E}[\boldsymbol{q}_{0}^{\ast }%
\boldsymbol{x}^{\ast \prime }]\right) ^{-1} \mathbb{E}[\boldsymbol{q}_{0}^{\ast
}y]. \end{equation*}
We know that, under Assumptions \ref{assumption kernel}%
--\ref{assumption Gbeta}, $\boldsymbol{\hat{\beta}}\xrightarrow[]{p}%
\boldsymbol{\beta }_{0}$, and, under Assumptions \ref{assumption kernel}--%
\ref {assumption moments Y} and \ref{assumption G*beta}, $\boldsymbol{\hat{%
\beta}}^{\ast }\xrightarrow[]{p}\boldsymbol{\beta }_{0}^{\ast }$, because
the sample moment conditions (\ref{sample mom cond IR}) and (\ref{sample mom
cond CF}) converge uniformly to, respectively, the moment conditions (\ref{%
mom cond IR}) and (\ref{mom cond CF}). We now show that $\alpha _{0}^{\ast
}=\alpha _{0}$ and $\boldsymbol{\gamma }_{0}^{\ast }=\boldsymbol{\gamma }%
_{0} $. Since $\boldsymbol{\beta }_{0}$ and $\boldsymbol{\beta }_{0}^{\ast }$
are unique, it suffices to show that $(\alpha _{0}^{\ast },\boldsymbol{%
\gamma }_{0}^{\ast })$ satisfies the moment condition of the
instrument-residual approach. Letting $\varepsilon _{0}^{\ast }\coloneqq
y-\alpha _{0}^{\ast }t-\boldsymbol{r}^{\prime }\boldsymbol{\gamma }%
_{0}^{\ast }-\phi _{0}\zeta _{0}(\boldsymbol{c})$, the first component of (%
\ref{mom cond IR}) evaluated at $(\alpha _{0}^{\ast },\boldsymbol{\gamma }%
_{0}^{\ast })$ is
\begin{eqnarray*} \mathbb{E}\left[ z^{\ast }(\zeta
_{0})\left( y-\alpha _{0}^{\ast }t- \boldsymbol{r}^{\prime }\boldsymbol{%
\gamma }_{0}^{\ast }\right) \right] &=& \mathbb{E}\left[ z^{\ast }(\zeta
_{0})\varepsilon _{0}^{\ast }\right] +\phi _{0}\mathbb{E}\left[ z^{\ast
}(\zeta _{0})\zeta _{0}(\boldsymbol{c})\right] \\ &=&\mathbb{E}\left[
z^{\ast }(\zeta _{0})\varepsilon _{0}^{\ast }\right] = \mathbb{E}\left[
z\varepsilon _{0}^{\ast }\right] -\mathbb{E}\left[ \zeta _{0}(\boldsymbol{c}%
)\varepsilon _{0}^{\ast }\right] , 
\end{eqnarray*}
where the second equality
uses the fact that $z-\mathbb{E}[z\mid \boldsymbol{c}]$ is orthogonal to any
function of $\boldsymbol{c}$. Since both $\mathbb{E}\left[ z\varepsilon
_{0}^{\ast }\right] $ and $\mathbb{E}\left[ \zeta _{0}(\boldsymbol{c}%
)\varepsilon _{0}^{\ast }\right] $ are zero by (\ref{mom cond CF}), it
follows that $(\alpha _{0}^{\ast },\boldsymbol{\gamma }_{0}^{\ast }) $
satisfies (\ref{mom cond IR}). The orthogonality condition $\mathbb{E}\left[
z^{\ast }(\zeta _{0})\left( y-\alpha _{0}t-\boldsymbol{r}^{\prime }%
\boldsymbol{\gamma }_{0}\right) \right] =0$ also provides an explicit
expression for $\alpha _{0}=\alpha _{0}^{\ast }$. Expanding the expectation,%
\begin{equation*} \mathbb{E}\left[ z^{\ast }(\zeta _{0})y\right] -\alpha _{0}%
\mathbb{E}\left[ z^{\ast }(\zeta _{0})t\right] -\mathbb{E}\left[ z^{\ast
}(\zeta _{0}) \boldsymbol{r}^{\prime }\right] \boldsymbol{\gamma }_{0}=0. 
\end{equation*} 
Since $\mathbb{E}\left[ z^{\ast }(\zeta _{0})\boldsymbol{r}%
^{\prime }\right] =\mathbf{0}_{k}^{\prime }$ by construction, it follows that%
\begin{equation*} \alpha _{0}=\frac{\mathbb{E}\left[ z^{\ast }(\zeta
_{0})y\right] }{\mathbb{E} \left[ z^{\ast }(\zeta _{0})t\right] }=\frac{%
\mathbb{C}\text{ov}(z^{\ast }(\zeta _{0})y)}{\mathbb{C}\text{ov}(z^{\ast
}(\zeta _{0})t)}, 
\end{equation*} 
which matches $\alpha_{\text{rich}}$ in equation (\ref{alpharich}).
\section{\label{App Neyman}Neyman
Orthogonality} Let $\boldsymbol{\theta }$ be a target (finite-dimensional)\
parameter and $\eta \in \mathcal{H}$ a (potentially high-dimensional or
infinite-dimensional) nuisance parameter. Consider a moment (or score)
function $\boldsymbol{m}(\boldsymbol{w},\boldsymbol{\theta },\eta )$ such
that $\mathbb{E}\left[ \boldsymbol{m}(\boldsymbol{w},\boldsymbol{\theta }%
_{0},\eta _{0})\right] =\boldsymbol{0}_{\operatorname{dim}(\boldsymbol{\theta })}$, where $\boldsymbol{w}$ denotes
the data. We say that the moment function $\boldsymbol{m}(\boldsymbol{w},%
\boldsymbol{\theta },\eta )$ is Neyman orthogonal at $(\boldsymbol{\theta }%
_{0},\eta _{0})$ with respect to $\eta $ if the Gateaux derivative \begin{%
equation*} \left. \frac{d}{dr}\mathbb{E}\left[ \boldsymbol{m}(\boldsymbol{w}%
, \boldsymbol{\theta }_{0},\eta _{0}+rh)\right] \right\vert _{r=0} \end{%
equation*}is zero for all directions $h\in \mathcal{H}$ \citep[]{chernozhukov2018}. Intuitively, Neyman orthogonality means that the moment
condition is locally insensitive to small perturbations of $\eta $ around $%
\eta _{0}$. 
\paragraph{Instrument-residual approach.} For the moment
condition (\ref{mom cond IR}), the Gateaux derivative with respect to $\zeta 
$ is
\begin{equation*} \left. \frac{d}{dr}\mathbb{E}\left[ \boldsymbol{q}%
\big(\zeta _{0}(\boldsymbol{c} )+rh(\boldsymbol{c})\big)(y-\boldsymbol{x}^{\prime }%
\boldsymbol{\beta }_{0}) \right] \right\vert _{r=0}=-\mathbb{E}\left[ h(%
\boldsymbol{c})(y-\boldsymbol{\ x}^{\prime }\boldsymbol{\beta }_{0})%
\boldsymbol{e}_{1,k+1}\right] , 
\end{equation*}
where $\boldsymbol{e}%
_{1,k+1}\coloneqq(1,\boldsymbol{0}_{k}^{\prime })^{\prime }$. Thus, the
moment function of the instrument-residual approach is Neyman orthogonal if
and only if $\mathbb{E}\left[ h(\boldsymbol{c})(y-\boldsymbol{x}^{\prime }%
\boldsymbol{\beta }_{0})\right] =0$ for all directions $h(\boldsymbol{c})$,
and hence, by the law of iterated expectations, if and only if the correct
specification condition $\mathbb{E}\left[ y-\boldsymbol{x}^{\prime }%
\boldsymbol{\beta }_{0}\mid \boldsymbol{c}\right] =0$ holds. 
\paragraph{%
Control-function approach.} For the moment condition (\ref{mom cond CF}),
the Gateaux derivative with respect to $\zeta $ is 
\begin{eqnarray} &&\left. 
\frac{d}{dr}\mathbb{E}\Bigl[ \boldsymbol{q}^{\ast }(\zeta _{0}( \boldsymbol{c%
})+rh(\boldsymbol{c}))\bigl\{ y-\alpha t-\boldsymbol{x}^{\prime } 
\boldsymbol{\gamma }_{0}-\phi _{0}\left( \zeta _{0}(\boldsymbol{c})+rh( 
\boldsymbol{c})\right) \bigr\} \Bigr] \right\vert _{r=0} \notag \\ &=&\left. 
\mathbb{E}\Bigl[ h(\boldsymbol{c})\bigl\{y-\alpha t-\boldsymbol{x} ^{\prime }%
\boldsymbol{\gamma }_{0}-\phi _{0}\left( \zeta _{0}(\boldsymbol{c} )+rh(%
\boldsymbol{c})\right) \bigr\}\boldsymbol{e}_{1,k+2}-\boldsymbol{q}^{\ast
}(\zeta _{0}(\boldsymbol{c})+rh(\boldsymbol{c}))\phi _{0}h(\boldsymbol{c})
\Bigr] \right\vert _{r=0} \notag \\ &=&\mathbb{E}\Bigl[ h(\boldsymbol{c})(y-%
\boldsymbol{x}^{\ast \prime } \boldsymbol{\beta }_{0}^{\ast })\boldsymbol{e}%
_{1,k+2}-\boldsymbol{q} _{0}^{\ast }\phi _{0}h(\boldsymbol{c})\Bigr] \notag
\\ &=& \begin{pmatrix} \mathbb{E}\left[ h(\boldsymbol{c})(y-\boldsymbol{x}%
^{\ast \prime } \boldsymbol{\beta }_{0}^{\ast })\right] -\phi _{0}\mathbb{E}%
\left[ h( \boldsymbol{c})\zeta _{0}(\boldsymbol{c})\right]\\ -\phi _{0}%
\mathbb{E}\left[ h(\boldsymbol{c})z\right] \\ -\phi _{0}\mathbb{E}\left[ h(%
\boldsymbol{c})\boldsymbol{c}\right]\end{pmatrix} . \label{Gateaux CF} \end{%
eqnarray} 
Clearly, (\ref{Gateaux CF}) equals $\boldsymbol{0}_{k+2}$ for all
directions $h(\boldsymbol{c})$ only under very restrictive conditions (e.g., 
$\mathbb{E}\left[ y-\boldsymbol{x}^{\ast \prime }\boldsymbol{\beta }%
_{0}^{\ast }\mid \boldsymbol{c}\right] =0$ and $\phi _{0}=0$), meaning that the
moment function of the control-function approach is not Neyman orthogonal,
except for very specific cases. 

\section{\label{App proofs}Proofs} It is
convenient to reparametrize $\zeta (\boldsymbol{c})$ as $\zeta (\boldsymbol{c})=\theta _{2}(%
\boldsymbol{c})/\theta _{1}(\boldsymbol{c})$, where $\theta _{1}(\boldsymbol{%
	c})\coloneqq f(\boldsymbol{c})$, the density of $\boldsymbol{c}$, and $%
\theta _{2}(\boldsymbol{c})\coloneqq\mathbb{E}\left[z\mid \boldsymbol{c}\right]f(%
\boldsymbol{c})$. Correspondingly, we estimate the $2\times 1$ vector $%
\boldsymbol{\theta }(\boldsymbol{c})\coloneqq(\theta _{1}(\boldsymbol{c}%
),\theta _{2}(\boldsymbol{c}))^{\prime }$ by 
\begin{equation*}
	\boldsymbol{\hat{\theta}}(\boldsymbol{c})\coloneqq\frac{1}{n}%
	\sum_{i=1}^{n}K_{h}(\boldsymbol{c}-\boldsymbol{c}_{i})\boldsymbol{\tilde{z}}%
	_{i},
\end{equation*}%
where $\boldsymbol{\tilde{z}}_{i}\coloneqq(1,z_{i})^{\prime }$ (so that $%
\hat{\theta}_{1}(\boldsymbol{c})=n^{-1}\sum_{i=1}^{n}K_{h}(\boldsymbol{c}-%
\boldsymbol{c}_{i})$, and $\hat{\theta}_{2}(\boldsymbol{c}%
)=n^{-1}\sum_{i=1}^{n}K_{h}(\boldsymbol{c}-\boldsymbol{c}_{i})z_{i}$). We
will use the following notation. For a matrix $\boldsymbol{A}\in \mathbb{R}^{m\times
n}$ with generic element $A_{ij}$, $\Vert \boldsymbol{A}\Vert $ denotes the
Frobenius norm $\sqrt{\sum_{i=1}^{m}\sum_{j=1}^{n}\left(A_{ij}\right)^{2}}$, and for a
matrix valued function $\boldsymbol{A}:\mathcal{X\subseteq }\mathbb{R}%
^{p}\mapsto \mathbb{R}^{m\times n}$, $\Vert \boldsymbol{A}\Vert _{\infty }$
denotes the supremum norm $\sup_{\boldsymbol{x}\in \mathcal{X}}\left\Vert 
\boldsymbol{A}(\boldsymbol{x})\right\Vert $. Finally, let $\boldsymbol{w}\coloneqq(y,\boldsymbol{x%
}^{\prime },z)^{\prime }$ and $\boldsymbol{w}^{\ast }\coloneqq(y,\boldsymbol{x}^{\ast \prime
},z)^{\prime }$. 

\begin{pff} [Proof of Theorem 
\ref{Th asy distr}] 
The theorem can be proved using Theorem 8.12 in \cite{NMcF1994}
(henceforth NM), which in turn relies on Theorem 8.11 in NM. The partial derivatives of the moment function $
\boldsymbol{g}(\boldsymbol{w},\boldsymbol{\beta },\zeta )\coloneqq%
\boldsymbol{q}(\zeta )(y-\boldsymbol{x}^{\prime }\boldsymbol{\beta })$ with
respect to $\boldsymbol{\beta }$ and $\boldsymbol{\theta }$ are\footnote{%
	Here, differentiation is performed with respect to the pointwise evaluation $%
	\boldsymbol{\theta }(\boldsymbol{c})$, not with respect to the function $%
	\boldsymbol{c}\mapsto \boldsymbol{\theta }$ as a whole.} 
\begin{equation}
	\underset{\left( k+1\right) \times \left( k+1\right) }{\underbrace{%
			\boldsymbol{H}_{\boldsymbol{\beta }}(\boldsymbol{w},\boldsymbol{\beta },%
			\boldsymbol{\theta })}}\coloneqq\frac{\partial \boldsymbol{g}(\boldsymbol{w},%
		\boldsymbol{\beta },\zeta )}{\partial \boldsymbol{\beta }^{\prime }}=-%
	\boldsymbol{q}(\zeta )\boldsymbol{x}^{\prime }  \label{Hbeta}
\end{equation}%
and 
\begin{equation}
	\underset{\left( k+1\right) \times 2}{\underbrace{\boldsymbol{H}_{%
				\boldsymbol{\theta }}(\boldsymbol{w},\boldsymbol{\beta },\boldsymbol{\theta }%
			)}}\coloneqq\frac{\partial \boldsymbol{g}(\boldsymbol{w},\boldsymbol{\beta }%
		,\zeta )}{\partial \boldsymbol{\theta }^{\prime }}=\frac{y-\boldsymbol{x}%
		^{\prime }\boldsymbol{\beta }}{\theta _{1}}\boldsymbol{e}_{1,k+1}\left(
	\zeta ,-1\right) ,  \label{Hgamma}
\end{equation}%
where $\boldsymbol{e}_{1,k+1}\coloneqq(1,\boldsymbol{0}_{k}^{\prime
})^{\prime }$. Also, we let $\boldsymbol{g}(\boldsymbol{w},\boldsymbol{\theta })%
\coloneqq\boldsymbol{g}(\boldsymbol{w},\boldsymbol{\beta }_{0},\boldsymbol{%
	\theta })$, $\boldsymbol{H}_{\boldsymbol{\beta }}(\boldsymbol{w},\boldsymbol{%
	\theta })\coloneqq\boldsymbol{H}_{\boldsymbol{\beta }}(\boldsymbol{w},%
\boldsymbol{\beta }_{0},\boldsymbol{\theta })$ and $\boldsymbol{H}_{%
	\boldsymbol{\theta }}(\boldsymbol{w},\boldsymbol{\theta })\coloneqq%
\boldsymbol{H}_{\boldsymbol{\theta }}(\boldsymbol{w},\boldsymbol{\beta }_{0},%
\boldsymbol{\theta })$. 

We now verify the conditions of Theorem 8.11 in NM. Our Assumptions \ref%
{assumption kernel} and \ref{assumption smoothness} correspond to Assumption
8.1 and 8.2 in NM. Assumption 8.3 in NM is satisfied since $%
\mathbb{E}\left[\left\Vert \boldsymbol{\tilde{z}}\right\Vert ^{p}\right]$ (and hence $%
\mathbb{E}\left[\left\Vert \boldsymbol{\tilde{z}}\right\Vert ^{p}\mid \boldsymbol{c}\right]$) is finite for any $p>0$, due to the fact that the instrument $z$ is
binary, and $f_{0}(\boldsymbol{c})$ is bounded from above by Assumption \ref%
{assumption c}. Theorem 8.11 in NM also requires the following conditions:

\begin{enumerate}
	\item[(a)] there is a vector of functionals $\boldsymbol{l}(\boldsymbol{w},%
	\boldsymbol{\theta })$ that is linear in $\boldsymbol{\theta }$ such that
	for $\left\Vert \boldsymbol{\theta }-\boldsymbol{\theta }_{0}\right\Vert
	_{\infty }$ small enough and for some function $b(\boldsymbol{w})$ with $%
	\mathbb{E}\left[b(\boldsymbol{w})\right]<\infty $, 
	\begin{equation}
		\left\Vert \boldsymbol{g}(\boldsymbol{w},\boldsymbol{\theta })-\boldsymbol{g}%
		(\boldsymbol{w},\boldsymbol{\theta }_{0})-\boldsymbol{l}(\boldsymbol{w},%
		\boldsymbol{\theta }-\boldsymbol{\theta }_{0})\right\Vert \leq b(\boldsymbol{%
			w})\left\Vert \boldsymbol{\theta }-\boldsymbol{\theta }_{0}\right\Vert
		_{\infty }^{2};
	\end{equation}
	
	\item[(b)] for some function $d(\boldsymbol{w})$ with $\mathbb{E}\left[d(%
	\boldsymbol{w})^{2})\right]\leq \infty $, $\left\Vert \boldsymbol{l}(\boldsymbol{w},%
	\boldsymbol{\theta })\right\Vert \leq d(\boldsymbol{w})\left\Vert 
	\boldsymbol{\theta }\right\Vert _{\infty }$;
	
	\item[(c)] there exists a matrix $\boldsymbol{v}(\boldsymbol{c})$ such that $\int 
	\boldsymbol{l}(\boldsymbol{w},\boldsymbol{\theta })\,\mathrm{d}F_{0}(%
	\boldsymbol{w})=\int \boldsymbol{v}(\boldsymbol{c})\boldsymbol{\theta }(\boldsymbol{c})\,%
	\mathrm{d}\boldsymbol{c}$ for all $\left\Vert \boldsymbol{\theta }%
	\right\Vert _{\infty }<\infty $;
	
	\item[(d)] $\boldsymbol{v}(\boldsymbol{c})$ is continuous almost everywhere, $\int
	\left\Vert \boldsymbol{v}(\boldsymbol{c})\right\Vert \,\mathrm{d}\boldsymbol{c}<\infty $,
	and there is $\epsilon >0$ such that $\mathbb{E}\left[\mathrm{sup}_{\left\Vert 
		\boldsymbol{\mathrm{v}}\right\Vert \leq \epsilon }\left\Vert \boldsymbol{v}(\boldsymbol{c}+\boldsymbol{\mathrm{\boldsymbol{v}}}%
	)\right\Vert ^{4}\right]<\infty $;
	
	\item[(e)] $\mathbb{E}\left[\left\Vert \boldsymbol{g}(\boldsymbol{w},\boldsymbol{%
		\theta }_{0})\right\Vert ^{2}\right]<\infty $.
\end{enumerate}

Conditions (a)--(e) are verified as follows.

\begin{enumerate}
	\item[(a)] Let $\boldsymbol{h}_{j}(\boldsymbol{w},\boldsymbol{\theta })%
	\coloneqq\frac{\partial g_{j}(\boldsymbol{w},\boldsymbol{\theta })}{\partial 
		\boldsymbol{\theta }^{\prime }}$ and define the $2\times 2$ matrices $%
	\boldsymbol{T}_{j}(\boldsymbol{w},\boldsymbol{\theta })\coloneqq\frac{%
		\partial \boldsymbol{h}_{j}(\boldsymbol{w},\boldsymbol{\theta })}{\partial 
		\boldsymbol{\theta }}$. A second-order mean value expansion of the $j$-th
	element, $j=1,\ldots ,k+1$, of $\boldsymbol{g}(\boldsymbol{w},\boldsymbol{%
		\theta })$ around $\boldsymbol{\theta }_{0}$ yields%
	\begin{equation*}
		g_{j}(\boldsymbol{w},\boldsymbol{\theta })=g_{j}(\boldsymbol{w},\boldsymbol{%
			\theta }_{0})+\boldsymbol{h}_{j}(\boldsymbol{w},\boldsymbol{\theta }_{0})(%
		\boldsymbol{\theta -\theta }_{0})+\frac{1}{2}(\boldsymbol{\theta -\theta }%
		_{0})^{\prime }\boldsymbol{T}_{j}(\boldsymbol{w},\boldsymbol{\bar{\theta}})(%
		\boldsymbol{\theta -\theta }_{0}),
	\end{equation*}%
	where $\boldsymbol{\bar{\theta}}$ is a mean value, and therefore%
	\begin{eqnarray}
		\left\vert g_{j}(\boldsymbol{w},\boldsymbol{\theta })-g_{j}(\boldsymbol{w},%
		\boldsymbol{\theta }_{0})-\boldsymbol{h}_{j}(\boldsymbol{w},\boldsymbol{%
			\theta }_{0})(\boldsymbol{\theta -\theta }_{0})\right\vert &=&\frac{1}{2}%
		\left\vert (\boldsymbol{\theta -\theta }_{0})^{\prime }\boldsymbol{T}_{j}(%
		\boldsymbol{w},\boldsymbol{\bar{\theta}})(\boldsymbol{\theta -\theta }%
		_{0})\right\vert  \notag \\
		&\leq &\frac{1}{2}\left\Vert \boldsymbol{T}_{j}(\boldsymbol{w},\boldsymbol{%
			\bar{\theta}})\right\Vert \left\Vert \boldsymbol{\theta -\theta }%
		_{0}\right\Vert _{\infty }^{2},  \label{c bound}
	\end{eqnarray}%
	by the Cauchy-Schwarz inequality. Noting that the $\boldsymbol{h}_{j}(%
	\boldsymbol{w},\boldsymbol{\theta })$s are the $1\times 2$ rows of $%
	\boldsymbol{H}_{\boldsymbol{\theta }}(\boldsymbol{w},\boldsymbol{\theta })$,
	the componentwise bound (\ref{c bound}) implies 
	\begin{equation*}
		\left\Vert \boldsymbol{g}(\boldsymbol{w},\boldsymbol{\theta })-\boldsymbol{g}%
		(\boldsymbol{w},\boldsymbol{\theta }_{0})-\boldsymbol{H}_{\boldsymbol{\theta 
		}}(\boldsymbol{w},\boldsymbol{\theta }_{0})(\boldsymbol{\theta -\theta }%
		_{0})\right\Vert \leq \frac{1}{2}\left\Vert \left( \boldsymbol{T}_{1}(%
		\boldsymbol{w},\boldsymbol{\bar{\theta}}),\ldots ,\boldsymbol{T}_{k+1}(%
		\boldsymbol{w},\boldsymbol{\bar{\theta}})\right) \right\Vert \left\Vert 
		\boldsymbol{\theta -\theta }_{0}\right\Vert _{\infty }^{2}.
	\end{equation*}%
	Observe now that $\boldsymbol{T}_{2}(\boldsymbol{w},\boldsymbol{\bar{\theta}}%
	)=\ldots =\boldsymbol{T}_{k}(\boldsymbol{w},\boldsymbol{\bar{\theta}})=%
	\boldsymbol{0}_{2,2}$, and take 
	\begin{equation}
		\boldsymbol{l}(\boldsymbol{w},\boldsymbol{\theta })=\boldsymbol{H}_{%
			\boldsymbol{\theta }}(\boldsymbol{w},\boldsymbol{\theta }_{0})\boldsymbol{%
			\theta }  \label{l}
	\end{equation}%
	and $b(\boldsymbol{w})=\frac{1}{2}\left\Vert \boldsymbol{T}_{1}(\boldsymbol{w%
	},\boldsymbol{\bar{\theta}})\right\Vert $. Since $\boldsymbol{T}_{1}(%
	\boldsymbol{w},\boldsymbol{\theta })=\left( y-\boldsymbol{x}^{\prime }%
	\boldsymbol{\beta }\right) \boldsymbol{A}$, with 
	\begin{equation*}
		\boldsymbol{A}\coloneqq\frac{1}{\theta _{1}^{2}}%
		\begin{pmatrix}
			-2\zeta & 1 \\ 
			1 & 0%
		\end{pmatrix}%
		,
	\end{equation*}%
	it follows that $\mathbb{E}\left[b(\boldsymbol{w})\right]=\mathbb{E}\left[\left( y-%
	\boldsymbol{x}^{\prime }\boldsymbol{\beta }\right) \left\Vert \boldsymbol{A}\right\Vert \right]=%
	\mathbb{E}\left[\left\vert y-\boldsymbol{x}^{\prime }\boldsymbol{\beta }%
	\right\vert \right]\left\Vert \boldsymbol{A}\right\Vert ,$ and therefore that $\mathbb{E}\left[b(%
	\boldsymbol{w})\right]<\infty $ under Assumption \ref{assumption moments Y}.
	
	\item[(b)] From equation (\ref{l}), $\left\Vert \boldsymbol{l}(\boldsymbol{w}%
	,\boldsymbol{\theta })\right\Vert \leq \left\Vert \boldsymbol{H}_{%
		\boldsymbol{\theta }}(\boldsymbol{w},\boldsymbol{\theta }_{0})\right\Vert
	\left\Vert \boldsymbol{\theta }\right\Vert _{\infty }$, so take $d(%
	\boldsymbol{w})=\left\Vert \boldsymbol{H}_{\boldsymbol{\theta }}(\boldsymbol{%
		w},\boldsymbol{\theta }_{0})\right\Vert $. Since all entries of $\boldsymbol{%
		H}_{\boldsymbol{\theta }}(\boldsymbol{w},\boldsymbol{\theta }_{0})$ are
	bounded, $\mathbb{E}\left[d(\boldsymbol{w})^{2}\right]=\mathbb{E}\left[\left\Vert 
	\boldsymbol{H}_{\boldsymbol{\theta }}(\boldsymbol{w},\boldsymbol{\theta }%
	_{0})\right\Vert ^{2}\right]\leq \infty $.
	
	\item[(c)] Using again equation (\ref{l}), 
	\begin{eqnarray}
		\int \boldsymbol{l}(\boldsymbol{w},\boldsymbol{\theta })\,\mathrm{d}F_{0}(%
		\boldsymbol{w}) &=&\int \boldsymbol{H}_{\boldsymbol{\theta }}(\boldsymbol{w},%
		\boldsymbol{\theta }_{0})\boldsymbol{\theta }(\boldsymbol{c})\,f_{0}(%
		\boldsymbol{w})\,\mathrm{d}\boldsymbol{w}  \notag \\
		&=&\int \boldsymbol{H}_{\boldsymbol{\theta }}(\boldsymbol{w},\boldsymbol{%
			\theta }_{0})f_{0}(\boldsymbol{w}\mid \boldsymbol{c})f_{0}(\boldsymbol{c})%
		\boldsymbol{\theta }(\boldsymbol{c})\,\mathrm{d}\boldsymbol{w}  \notag \\
		&=&\int \mathbb{E}\left[ \boldsymbol{H}_{\boldsymbol{\theta }}(\boldsymbol{w}%
		,\boldsymbol{\theta }_{0})\mid \boldsymbol{c}\right] f_{0}(\boldsymbol{c})%
		\boldsymbol{\theta }(\boldsymbol{c})\,\mathrm{d}\boldsymbol{c}.  \label{LHS}
	\end{eqnarray}%
	Equating (\ref{LHS}) to $\int \boldsymbol{v}(\boldsymbol{c})\boldsymbol{%
		\theta }(\boldsymbol{c})\,\mathrm{d}\boldsymbol{c}$, and inserting
	expression (\ref{Hgamma}), we obtain 
	\begin{eqnarray}
		\underset{\left( k+1\right) \times 2}{\underbrace{\boldsymbol{v}(\boldsymbol{%
					c})}} &=&\mathbb{E}\left[ \boldsymbol{H}_{\boldsymbol{\theta }}(\boldsymbol{w%
		},\boldsymbol{\theta }_{0})\mid \boldsymbol{c}\right] f_{0}(\boldsymbol{c}) 
		\notag \\
		&=&\mathbb{E}\left[ \varepsilon \mid \boldsymbol{c}\right] \boldsymbol{e}%
		_{1,k+1}\left( \zeta _{0}(\boldsymbol{c}),-1\right) .  \label{v}
	\end{eqnarray}
	
	\item[(d)] The integral $\int \left\Vert \boldsymbol{v}(\boldsymbol{c}%
	)\right\Vert \,\mathrm{d}\boldsymbol{c\ }$is finite because $\boldsymbol{v}(%
	\boldsymbol{c})$ is continuous almost everywhere by Assumption \ref%
	{assumption smoothness}, is bounded on $\mathcal{C}$, and $\mathcal{C}$ is
	compact. Boundness of $\boldsymbol{v}(\boldsymbol{c})$ also implies that
	there is $\epsilon >0$ such that $\mathbb{E}\left[\mathrm{sup}_{\left\Vert 
		\boldsymbol{\mathrm{v}}\right\Vert \leq \epsilon }\left\Vert \boldsymbol{v}(\boldsymbol{c}%
	+\boldsymbol{\mathrm{v}})\right\Vert ^{4}\right]<\infty $.
	
	\item[(e)] Since $\mathbb{E}\left[\left\Vert \boldsymbol{g}(\boldsymbol{w},%
	\boldsymbol{\theta }_{0})\right\Vert ^{2}\right]=\mathbb{E}%
	\left[\sum_{j=1}^{k+1}g_{j}^{2}(\boldsymbol{w},\boldsymbol{\theta }_{0})\right]$, it
	follows that $\mathbb{E}\left[\left\Vert \boldsymbol{g}(\boldsymbol{w},%
	\boldsymbol{\theta }_{0})\right\Vert ^{2}\right]<\infty $ if and only if $\mathbb{E%
	}\left[g_{j}^{2}(\boldsymbol{w},\boldsymbol{\theta }_{0})\right]<\infty $ for each $%
	j=1,\ldots ,k+1$. By the Cauchy-Schwarz inequality $\mathbb{E}\left[g_{j}^{2}(%
	\boldsymbol{w},\boldsymbol{\theta }_{0})\right]=\mathbb{E}\left[q{_{j}^{2}}(\boldsymbol{%
		\theta }_{0})(y-\boldsymbol{x}^{\prime }\boldsymbol{\beta }_{0})^{2}\right]\leq 
	\sqrt{\mathbb{E}\left[q{_{j}^{4}}(\boldsymbol{\theta }_{0})\right]\mathbb{E}\left[(y-%
		\boldsymbol{x}^{\prime }\boldsymbol{\beta }_{0})^{4}\right]},$ which, under
	Assumption \ref{assumption moments Y}, is finite for each $j=2,\ldots ,k+1$.
	For $j=1$, note that $\mathbb{E}\left[q{_{1}^{4}}(\boldsymbol{\theta }_{0})\right]=%
	\mathbb{E}\left[(z^{\ast }){^{4}}\right]<\infty $, because $\mathbb{E}\left[(z^{\ast }){^{4}}%
	\right]<\mathbb{E}\left[z^{4}\right]$, due to Jensen's inequality and the law of total
	expectation, and $\mathbb{E}\left[z{^{4}}\right]<\infty $ since $z$ is binary.
\end{enumerate}

So far we have showed that Theorem 8.11 in NM applies to our setting.
Theorem 8.12 in NM requires the following additional conditions:

\begin{enumerate}
	\item[(f)] $\boldsymbol{\hat{\beta}}\xrightarrow[]{p}\boldsymbol{\beta }_{0}$%
	;
	
	\item[(g)] for $\Vert \boldsymbol{\theta }-\boldsymbol{\theta }_{0}\Vert
	_{\infty }$ sufficiently small, $\boldsymbol{g}(\boldsymbol{w},\boldsymbol{%
		\beta },\boldsymbol{\theta })$ is continuously differentiable in $%
	\boldsymbol{\beta }$ on a neighborhood $\mathcal{N}$ of $\boldsymbol{\beta }%
	_{0}$, and there exists a function $f(\boldsymbol{w})$ with $\mathbb{E}\left[f(%
	\boldsymbol{w})\right]<\infty $ and a constant $\varepsilon >0$ such that $\Vert 
	\boldsymbol{H}_{\boldsymbol{\beta }}(\boldsymbol{w},\boldsymbol{\beta },%
	\boldsymbol{\theta })-\boldsymbol{H}_{\boldsymbol{\beta }}(\boldsymbol{w},%
	\boldsymbol{\beta }_{0},\boldsymbol{\theta }_{0})\Vert \leq f(\boldsymbol{w})%
	\big(\Vert \boldsymbol{\beta }-\boldsymbol{\beta }_{0}\Vert ^{\varepsilon
	}+\Vert \boldsymbol{\theta }-\boldsymbol{\theta }_{0}\Vert _{\infty
	}^{\varepsilon }\big)$;
	
	\item[(h)] $\mathbb{E}[\boldsymbol{H}_{\boldsymbol{\beta }}(\boldsymbol{w},%
	\boldsymbol{\theta }_{0})]$ exists and is nonsingular.
	
	Conditions (f)--(h) are verified next.
	
	\item[(f)] Under Assumptions \ref{assumption kernel}--\ref{assumption smoothness}, the Nadaraya-Watson estimator $\hat{\zeta}(\boldsymbol{c}_{i})$ is
	uniformly consistent. Adding Assumption \ref{assumption Gbeta}, we also know
	that 
	\begin{equation*}
		\boldsymbol{\hat{\beta}}(\zeta _{0})\coloneqq\left( \sum_{i=1}^{n}\boldsymbol{q}{_{i}}%
		(\zeta _{0})\boldsymbol{x}_{i}^{\prime }\right) ^{-1}\left( \sum_{i=1}^{n}\boldsymbol{q}{%
			_{i}}(\zeta _{0})y_{i}\right) \xrightarrow[]{p}\boldsymbol{\beta }_{0}.
	\end{equation*}%
	Consistency of the two-step estimator $\boldsymbol{\hat{\beta}}$ then
	follows from continuity of the map from $\hat{\zeta}$ to $\boldsymbol{\hat{%
			\beta}}$ and Slutsky's theorem.
	
	\item[(g)] From equation (\ref{Hbeta}), 
	\begin{equation*}
		\boldsymbol{H}_{\boldsymbol{\beta }}(\boldsymbol{w},\boldsymbol{\beta },%
		\boldsymbol{\theta })-\boldsymbol{H}_{\boldsymbol{\beta }}(\boldsymbol{w},%
		\boldsymbol{\beta }_{0},\boldsymbol{\theta }_{0})=-\left( \boldsymbol{q}(%
		\boldsymbol{\theta })-\boldsymbol{q}(\boldsymbol{\theta }_{0})\right) 
		\boldsymbol{x}^{\prime }.
	\end{equation*}%
	Thus, 
	\begin{equation*}
		\left\Vert \boldsymbol{H}_{\boldsymbol{\beta }}(\boldsymbol{w},\boldsymbol{%
			\beta },\boldsymbol{\theta })-\boldsymbol{H}_{\boldsymbol{\beta }}(%
		\boldsymbol{w},\boldsymbol{\beta }_{0},\boldsymbol{\theta }_{0})\right\Vert
		=\left\Vert \left( \boldsymbol{q}(\boldsymbol{\theta })-\boldsymbol{q}(%
		\boldsymbol{\theta }_{0})\right) \boldsymbol{x}^{\prime }\right\Vert \leq
		\left\Vert \boldsymbol{q}(\boldsymbol{\theta })-\boldsymbol{q}(\boldsymbol{%
			\theta }_{0})\right\Vert \left\Vert \boldsymbol{x}\right\Vert .
	\end{equation*}%
	Take $f(\boldsymbol{w})=\left\Vert \boldsymbol{x}\right\Vert $. Then $%
	\mathbb{E}\left[f(\boldsymbol{w})\right]<\infty $ by Assumption \ref{assumption moments Y}
	and the fact that $t$ is binary. We now show that, for $\Vert \boldsymbol{%
		\theta }-\boldsymbol{\theta }_{0}\Vert _{\infty }$ sufficiently small, $%
	\Vert \boldsymbol{H}_{\boldsymbol{\beta }}(\boldsymbol{w},\boldsymbol{\beta }%
	,\boldsymbol{\theta })-\boldsymbol{H}_{\boldsymbol{\beta }}(\boldsymbol{w},%
	\boldsymbol{\beta }_{0},\boldsymbol{\theta }_{0})\Vert \leq f(\boldsymbol{w}%
	)\Vert \boldsymbol{\theta }-\boldsymbol{\theta }_{0}\Vert _{\infty }$. Since 
	$\boldsymbol{q}(\boldsymbol{\theta })-\boldsymbol{q}(\boldsymbol{\theta }%
	_{0})=(-\zeta +\zeta _{0},\boldsymbol{0}_{k}^{\prime })^{\prime }$,%
	\begin{eqnarray}
		\left\Vert \boldsymbol{q}(\boldsymbol{\theta })-\boldsymbol{q}(\boldsymbol{%
			\theta }_{0})\right\Vert &=&\left\vert -\zeta +\zeta _{0}\right\vert  \notag
		\\
		&=&\left\vert \frac{\theta _{2}\theta _{1,0}-\theta _{2,0}\theta _{1}}{%
			\theta _{1}\theta _{1,0}}\right\vert .  \label{ratio}
	\end{eqnarray}
	
	The numerator of (\ref{ratio}) is $\theta _{2}\theta _{1,0}-\theta
	_{2,0}\theta _{1}=(\theta _{2}-\theta _{2,0})\theta _{1,0}+\theta
	_{2,0}(\theta _{1,0}-\theta _{1})$, so by the triangle inequality $%
	\left\vert \theta _{2}\theta _{1,0}-\theta _{2,0}\theta _{1}\right\vert \leq
	\left\vert (\theta _{2}-\theta _{2,0})\theta _{1,0}\right\vert +\left\vert
	\theta _{2,0}(\theta _{1,0}-\theta _{1})\right\vert \leq \left\vert (\theta
	_{2}-\theta _{2,0})\theta _{1,0}\right\vert +\left\vert \theta _{2,0}(\theta
	_{1,0}-\theta _{1})\right\vert .$ By Assumptions \ref{assumption smoothness}
	and \ref{assumption c}, $\left\vert \theta _{2}\theta _{1,0}-\theta
	_{2,0}\theta _{1}\right\vert \leq C\Vert \boldsymbol{\theta }-\boldsymbol{%
		\theta }_{0}\Vert _{\infty }$, where $C$ is a constant that depends on the
	smoothness of $\mathbb{E}\left[z\mid \boldsymbol{c}\right]$ and the bound on $f(\boldsymbol{%
		c})$. As for the denominator of (\ref{ratio}), by Assumption \ref{assumption
		c}, there exists a constant $\delta >0$ such that $\theta _{1}\theta
	_{1,0}\geq \delta $ for all $\boldsymbol{c}\in \mathcal{C}.$ Combining the
	bounds on numerator and denominator, we get 
	\begin{equation}
	\left\vert \zeta (\boldsymbol{c}%
	)-\zeta _{0}(\boldsymbol{c})\right\vert \leq \delta ^{-1}C\Vert \boldsymbol{%
		\theta }-\boldsymbol{\theta }_{0}\Vert _{\infty }.\label{delta}
	\end{equation}
	\item[(h)] Existence is guaranteed by Assumption \ref{assumption moments Y},
	nonsingularity by Assumption \ref{assumption Gbeta}.
\end{enumerate}

Having showed that Theorems 8.11 and 8.12 in NM apply to our context, to
complete the proof we only need to derive the expression for the $\left(
k+1\right) \times 1$ correction term $\boldsymbol{\varphi }(\boldsymbol{w})$%
. From (\ref{v}),%
\begin{align*}
	\boldsymbol{v}(\boldsymbol{c})\boldsymbol{\tilde{z}} &=\mathbb{E}\left[
	\varepsilon \mid \boldsymbol{c}\right] \boldsymbol{e}_{1,k+1}\left( \zeta _{0}(%
	\boldsymbol{c}),-1\right) 
	\begin{pmatrix}
		1 \\ 
		z%
	\end{pmatrix}
	\\
	&=\mathbb{E}\left[ \varepsilon \mid \boldsymbol{c}\right] \left( \zeta _{0}(%
	\boldsymbol{c})-z\right) \boldsymbol{e}_{1,k+1} \\
	&=-\mathbb{E}\left[ \varepsilon \mid \boldsymbol{c}\right] 
	\begin{pmatrix}
		z^{\ast }(\zeta _{0}) \\ 
		\boldsymbol{0}_{k}%
	\end{pmatrix}%
	,
\end{align*}%
and therefore $\boldsymbol{\varphi }(\boldsymbol{w})=(\varphi _{1}(%
\boldsymbol{w}),\boldsymbol{0}_{k}^{\prime })^{\prime }$ with%
\begin{equation*}
	\varphi _{1}(\boldsymbol{w})\coloneqq-\mathbb{E}\left[ \varepsilon \mid %
	\boldsymbol{c}\right] z^{\ast }(\zeta _{0})+\mathbb{E}\left[ \mathbb{E}%
	\left[ \varepsilon \mid \boldsymbol{c}\right] z^{\ast }(\zeta _{0})\right] =-%
	\mathbb{E}\left[ \varepsilon \mid \boldsymbol{c}\right] z^{\ast }(\zeta _{0}),
\end{equation*}%
since $\mathbb{E}\left[ \mathbb{E}\left[ \varepsilon \mid \boldsymbol{c}\right]
\left( z-\mathbb{E}\left[ z\mid \boldsymbol{c}\right] \right) \right] =0$.
\end{pff} 

\begin{pff}[Proof of Theorem \protect\ref{cons var IR}]
	The result follows from Theorem 8.13 in NM, which requires a regularity condition that can be broken down in the two following conditions:
	
	\begin{itemize}
		\item[(i)] there exists a function $b(\boldsymbol{w})$ and constant $\epsilon > 0$ such that $\mathbb{E}[b(\boldsymbol{w})^2] < \infty$, and for $\Vert \boldsymbol{\theta} - \boldsymbol{\theta}_0 \Vert_\infty$ small enough,
		$\Vert \boldsymbol{g}(\boldsymbol{w}, \boldsymbol{\beta}, \boldsymbol{\theta}) - \boldsymbol{g}(\boldsymbol{w}, \boldsymbol{\beta}_0, \boldsymbol{\theta}_0)\Vert 
		\leq b(\boldsymbol{w}) \left( \Vert\boldsymbol{\beta} - \boldsymbol{\beta}_0\Vert^\epsilon + \Vert\boldsymbol{\theta} - \boldsymbol{\theta}_0\Vert_\infty^\epsilon \right);$
		
		\item[(j)] there exists a function $d(\boldsymbol{w})$ and constant $\epsilon > 0$ such that $\mathbb{E}[d(\boldsymbol{w})^2] < \infty$, and for $\Vert \boldsymbol{\theta} - \boldsymbol{\theta}_0 \Vert_\infty$ small enough, $
		\Vert\boldsymbol{\varphi}(\boldsymbol{w}, \boldsymbol{\beta}, \boldsymbol{\theta}) - \boldsymbol{\varphi}(\boldsymbol{w}, \boldsymbol{\beta}_0, \boldsymbol{\theta}_0)\Vert
		\leq d(\boldsymbol{w}) \left( \Vert\boldsymbol{\beta} - \boldsymbol{\beta}_0\Vert^\epsilon + \Vert\boldsymbol{\theta} - \boldsymbol{\theta}_0\Vert_\infty^\epsilon \right).$
	\end{itemize}
	
	\vspace{1em}
	Conditions (i) and (j) are verified as follows.
	
	\begin{itemize}
		\item[(i)] Let
		\begin{align*}
			\boldsymbol{\Delta g}(\boldsymbol{w}) 
			&\coloneqq \boldsymbol{g}(\boldsymbol{w}, \boldsymbol{\beta}, \boldsymbol{\theta}) - \boldsymbol{g}(\boldsymbol{w}, \boldsymbol{\beta}_0, \boldsymbol{\theta}_0) \\
			&= \boldsymbol{q}(\boldsymbol{\theta}) (y - \boldsymbol{x}' \boldsymbol{\beta}) - \boldsymbol{q}(\boldsymbol{\theta}_0) (y - \boldsymbol{x}' \boldsymbol{\beta}_0) \\
			&= \boldsymbol{q}(\boldsymbol{\theta}) \boldsymbol{x}'(\boldsymbol{\beta}_0 - \boldsymbol{\beta}) + (y - \boldsymbol{x}' \boldsymbol{\beta}_0)(\boldsymbol{q}(\boldsymbol{\theta}) - \boldsymbol{q}(\boldsymbol{\theta}_0)).
		\end{align*}
		
		Taking norms and applying the triangle inequality gives
		\[
		\Vert \boldsymbol{\Delta g}(\boldsymbol{w}) \Vert 
		\leq \sup_{\boldsymbol{\theta} } \Bigl\{\Vert \boldsymbol{q}(\boldsymbol{\theta}) \Vert  \Vert \boldsymbol{x} \Vert  \Vert \boldsymbol{\beta} - \boldsymbol{\beta}_0 \Vert
		+ \vert y - \boldsymbol{x}' \boldsymbol{\beta}_0\vert  \Vert \boldsymbol{q}(\boldsymbol{\theta}) - \boldsymbol{q}(\boldsymbol{\theta}_0) \Vert \Bigr\}.
		\]
		
		We can then set
		\[
		b(\boldsymbol{w}) = \max \left\{ 
		\sup_{\boldsymbol{\theta} } \Vert  \boldsymbol{q}(\boldsymbol{\theta}) \Vert  \Vert \boldsymbol{x} \Vert,
		\ \delta^{-1} C  \vert y - \boldsymbol{x}' \boldsymbol{\beta}_0\vert 
		\right\},
		\]
		which satisfies $\mathbb{E}[b(\boldsymbol{w})^2] < \infty$ by Assumption \ref{assumption moments Y}, boundedness of $t$, $z$ and equation (\ref{ratio}) 
		that guarantees that $\Vert \boldsymbol q(\boldsymbol \theta)-\boldsymbol q(\boldsymbol \theta_0)\Vert$ is bounded.
		
		\vspace{1em}
		\item[(j)] Letting $\varepsilon(\boldsymbol{\beta}) \coloneqq y - \boldsymbol{x}' \boldsymbol{\beta}$,
		\[
		\varphi_1(\boldsymbol{w}, \boldsymbol{\beta}, \boldsymbol{\theta}) - \varphi_1(\boldsymbol{w}, \boldsymbol{\beta}_0, \boldsymbol{\theta}_0) 
		= -\mathbb{E}[\varepsilon(\boldsymbol{\beta}) \mid \boldsymbol{c}] z^*(\zeta) + \mathbb{E}[\varepsilon(\boldsymbol{\beta}_0) \mid \boldsymbol{c}] z^*(\zeta_0),
		\]
		
		and therefore, using equation (\ref{delta}),
		\begin{align*}
			\Vert \varphi_1(\boldsymbol{w}, \boldsymbol{\beta}, \boldsymbol{\theta}) - \varphi_1(\boldsymbol{w}, \boldsymbol{\beta}_0, \boldsymbol{\theta}_0) \Vert
			&\leq \vert \mathbb{E}[\varepsilon(\boldsymbol{\beta}) \mid \boldsymbol{c}] - \mathbb{E}[\varepsilon(\boldsymbol{\beta}_0) \mid \boldsymbol{c}] \vert  \,\vert z^*(\zeta) \vert \\
			&\quad + \vert \mathbb{E}[\varepsilon(\boldsymbol{\beta}_0) \mid \boldsymbol{c}] \vert\,  \vert \zeta(\boldsymbol{c}) - \zeta_0(\boldsymbol{c}) \vert \\
			&\leq \Vert \mathbb{E}[ \boldsymbol{x}  \mid \boldsymbol{c}]\Vert  \Vert \boldsymbol{\beta} - \boldsymbol{\beta}_0 \Vert  \sup_{\boldsymbol{\theta}} \vert z^*(\zeta) \vert \\
			&\quad
			+ \vert \mathbb{E}[ \varepsilon(\boldsymbol{\beta}_0)  \mid \boldsymbol{c}] \vert  \delta^{-1} C  \Vert \boldsymbol{\theta} - \boldsymbol{\theta}_0 \Vert_\infty.
		\end{align*}		
		Given this, set
		\[
		d(\boldsymbol{w}) \coloneqq \max \left\{ 
		\delta^{-1} C  \vert\mathbb{E}[ \varepsilon(\boldsymbol{\beta}_0)  \mid \boldsymbol{c}]\vert,
		\Vert \mathbb{E}[ \boldsymbol{x}  \mid \boldsymbol{c}]\Vert  \sup_{\boldsymbol{\theta}} \vert z^*(\zeta) \vert
		\right\}.
		\]
		Assumptions \ref{assumption c} and \ref{assumption moments Y}, along with the binary nature of $t$ and $z$, guarantee finiteness of $\mathbb{E}[d(\boldsymbol w)^2].$ 
		Finally, note that, for $j = 2, \dots, k+1$, boundedness is guaranteed because $\varphi_j(\boldsymbol{w}, \boldsymbol{\beta}, \boldsymbol{\theta}) = 0$.
	\end{itemize}	
\end{pff}

\begin{pff}[Proof of Theorem \protect\ref{Th asy distr CF}]
The partial derivative of the moment function 
	\begin{equation*}
		\boldsymbol{g}(\boldsymbol{w}^{\ast },\boldsymbol{\beta }^{\ast },\zeta )%
		\coloneqq\boldsymbol{q}{^{\ast }}(\zeta )(y-\boldsymbol{x}^{\ast \prime }%
		\boldsymbol{\beta }^{\ast })=%
		\begin{pmatrix}
			z \\ 
			\boldsymbol{r} \\ 
			\zeta%
		\end{pmatrix}%
		(y-\boldsymbol{x}^{\ast \prime }\boldsymbol{\beta }^{\ast }),
	\end{equation*}%
	with respect to $\boldsymbol{\beta }^{\ast }$ is 
	\begin{equation}
		\underset{\left( k+2\right) \times \left( k+2\right) }{\underbrace{%
				\boldsymbol{H}_{\boldsymbol{\beta }^{\ast }}(\boldsymbol{w}^{\ast },%
				\boldsymbol{\beta }^{\ast },\boldsymbol{\theta })}}\coloneqq\frac{\partial 
			\boldsymbol{g}(\boldsymbol{w}^{\ast },\boldsymbol{\beta }^{\ast },\zeta )}{%
			\partial \boldsymbol{\beta }^{\ast \prime }}=-\boldsymbol{q}^{\ast }(\zeta )%
		\boldsymbol{x}^{\ast \prime }.  \label{Hbeta_cf}
	\end{equation}%
	The partial derivatives with respect to $\theta _{1}$ and $\theta _{2}$ are%
	\begin{equation*}
		\frac{\partial \boldsymbol{g}(\boldsymbol{w}^{\ast },\boldsymbol{\beta }%
			^{\ast },\zeta )}{\partial \theta _{1}}=\frac{1}{\theta _{1}}\left\{ 
		\begin{pmatrix}
			\boldsymbol{0}_{k+1} \\ 
			-\frac{\theta _{2}}{\theta _{1}}%
		\end{pmatrix}%
		(y-\boldsymbol{x}^{\ast \prime }\boldsymbol{\beta }^{\ast })+%
		\begin{pmatrix}
			z \\ 
			\boldsymbol{r} \\ 
			\frac{\theta _{2}}{\theta _{1}}%
		\end{pmatrix}%
		\frac{\theta _{2}}{\theta _{1}}\phi \right\}
	\end{equation*}%
	and%
	\begin{equation*}
		\frac{\partial \boldsymbol{g}(\boldsymbol{w}^{\ast },\boldsymbol{\beta }%
			^{\ast },\zeta )}{\partial \theta _{2}}=\frac{1}{\theta _{1}}\left\{ 
		\begin{pmatrix}
			\boldsymbol{0}_{k+1} \\ 
			1%
		\end{pmatrix}%
		(y-\boldsymbol{x}^{\ast \prime }\boldsymbol{\beta }^{\ast })-%
		\begin{pmatrix}
			z \\ 
			\boldsymbol{r} \\ 
			\frac{\theta _{2}}{\theta _{1}}%
		\end{pmatrix}%
		\phi \right\} ,
	\end{equation*}%
	and therefore the partial derivative with respect to $\boldsymbol{\theta }$
	is%
	\begin{eqnarray}
		\underset{\left( k+2\right) \times 2}{\underbrace{\boldsymbol{H}_{%
					\boldsymbol{\theta }}(\boldsymbol{w}^{\ast },\boldsymbol{\beta }^{\ast },%
				\boldsymbol{\theta })}}\coloneqq\frac{\partial \boldsymbol{g}(\boldsymbol{w}%
			^{\ast },\boldsymbol{\beta }^{\ast },\zeta )}{\partial \boldsymbol{\theta }%
			^{\prime }} &=&\frac{1}{\theta _{1}}\bigl\{ \left( y-\boldsymbol{x}^{\ast
			\prime }\boldsymbol{\beta }^{\ast }\right) \boldsymbol{e}_{k+2,k+2}\left(
		-\zeta ,1\right) -\phi \boldsymbol{q}{^{\ast }}(\zeta )\left( -\zeta
		,1\right) \bigl\}  \notag \\
		&=&\frac{1}{\theta _{1}}\bigl\{ \left( y-\boldsymbol{x}^{\ast \prime }%
		\boldsymbol{\beta }^{\ast }\right) \boldsymbol{e}_{k+2,k+2}-\phi \boldsymbol{%
			q}{^{\ast }}(\zeta )\bigr\} \left( -\zeta ,1\right)  \notag \\
		&=&\frac{1}{\theta _{1}}\bigl( \varepsilon ^{\ast }\boldsymbol{e}%
		_{k+2,k+2}-\phi \boldsymbol{q}{^{\ast }}(\zeta )\bigr) \left( -\zeta
		,1\right) ,  \label{Hgamma_cf}
	\end{eqnarray}%
	where $\boldsymbol{e}_{k+2,k+2}\coloneqq(\boldsymbol{0}_{k+1}^{\prime
	},1)^{\prime }$. We let $\boldsymbol{g}(\boldsymbol{w}^{\ast },\boldsymbol{%
		\theta })\coloneqq\boldsymbol{g}(\boldsymbol{w}^{\ast },\boldsymbol{\beta }%
	_{0}^{\ast },\boldsymbol{\theta })$, $\boldsymbol{H}_{\boldsymbol{\beta }%
		^{\ast }}(\boldsymbol{w}^{\ast },\boldsymbol{\theta })\coloneqq\boldsymbol{H}%
	_{\boldsymbol{\beta }^{\ast }}(\boldsymbol{w}^{\ast },\boldsymbol{\beta }%
	_{0}^{\ast },\boldsymbol{\theta })$ and $\boldsymbol{H}_{\boldsymbol{\theta }%
	}(\boldsymbol{w}^{\ast },\boldsymbol{\theta })\coloneqq\boldsymbol{H}_{%
		\boldsymbol{\theta }}(\boldsymbol{w}^{\ast },\boldsymbol{\beta }_{0}^{\ast },%
	\boldsymbol{\theta })$. Similar to the proof of Theorem \ref{Th asy distr},
	we will use Theorem 8.11 and 8.12 from NM. Theorem 8.11 requires the following
	conditions:
	
	\begin{enumerate}
		\item[(a$^{\ast}$)] there is a vector of functionals $\boldsymbol{l}(%
		\boldsymbol{w^{\ast}}, \boldsymbol{\theta })$ that is linear in $\boldsymbol{%
			\theta }$ such that for $\left\Vert \boldsymbol{\theta }-\boldsymbol{\theta }%
		_{0}\right\Vert _{\infty }$ small enough and for some function $b(%
		\boldsymbol{w})$ with $\mathbb{E}\left[b(\boldsymbol{w^*})\right]<\infty $, 
		\begin{equation}
			\left\Vert \boldsymbol{g}(\boldsymbol{w^*},\boldsymbol{\theta })-\boldsymbol{%
				g} (\boldsymbol{w^*},\boldsymbol{\theta }_{0})-\boldsymbol{l}(\boldsymbol{w^*%
			}, \boldsymbol{\theta }-\boldsymbol{\theta }_{0})\right\Vert \leq b(%
			\boldsymbol{w^*})\left\Vert \boldsymbol{\theta }-\boldsymbol{\theta }%
			_{0}\right\Vert _{\infty }^{2};
		\end{equation}
		
		\item[(b$^{\ast}$)] for some function $d(\boldsymbol{w^*})$ with $\mathbb{E}%
		(d( \boldsymbol{w^*})^{2})\leq \infty $, $\left\Vert \boldsymbol{l}(%
		\boldsymbol{w^*}, \boldsymbol{\theta })\right\Vert \leq d(\boldsymbol{w^*}%
		)\left\Vert \boldsymbol{\theta }\right\Vert _{\infty }$;
		
		\item[(c$^{\ast}$)] there exists a matrix $\boldsymbol{v}^*(\boldsymbol{c})$ such that $%
		\int \boldsymbol{l}(\boldsymbol{w^*},\boldsymbol{\theta })\,\mathrm{d}F_{0}( 
		\boldsymbol{w^*})=\int \boldsymbol{v}^*(\boldsymbol{c})\boldsymbol{\theta }(\boldsymbol{c}%
		)\, \mathrm{d}\boldsymbol{c}$ for all $\left\Vert \boldsymbol{\theta }
		\right\Vert _{\infty }<\infty $;
		
		\item[(d$^{\ast}$)] $\boldsymbol{v}^*(\boldsymbol{c})$ is continuous almost everywhere, $%
		\int \left\Vert \boldsymbol{v}^*(\boldsymbol{c})\right\Vert \,\mathrm{d}\boldsymbol{c}%
		<\infty $, and there is $\epsilon >0$ such that $\mathbb{E}\left[\mathrm{sup}%
		_{\left\Vert \boldsymbol{\mathrm{v}}\right\Vert \leq \epsilon }\left\Vert \boldsymbol{v}^*(\boldsymbol{%
			c}+\boldsymbol{\mathrm{v}} )\right\Vert ^{4}\right]<\infty $;
		
		\item[(e$^{\ast}$)] $\mathbb{E}\left[\left\Vert \boldsymbol{g}(\boldsymbol{w^*},%
		\boldsymbol{\ \theta }_{0})\right\Vert ^{2}\right]<\infty $.
	\end{enumerate}
	
	Conditions (a$^{\ast}$)--(e$^{\ast}$) are verified as follows.
	
	\begin{enumerate}
		\item[(a$^{\ast}$)] Let $\boldsymbol{h}_{j}(\boldsymbol{w}^{\ast },%
		\boldsymbol{\theta })\coloneqq\frac{\partial g_{j}(\boldsymbol{w}^{\ast },%
			\boldsymbol{\theta })}{\partial \boldsymbol{\theta }^{\prime }}$ and define
		the $2\times 2$ matrices $\boldsymbol{T}_{j}(\boldsymbol{w}^{\ast },%
		\boldsymbol{\theta })\coloneqq\frac{\partial \boldsymbol{h}_{j}(\boldsymbol{w%
			}^{\ast },\boldsymbol{\theta })}{\partial \boldsymbol{\theta }}$. A
		second-order mean value expansion of the $j$-th element, $j=1,\ldots ,k+2$,
		of $\boldsymbol{g}(\boldsymbol{w}^{\ast },\boldsymbol{\theta })$ around $%
		\boldsymbol{\theta }_{0}$ yields%
		\begin{equation*}
			g_{j}(\boldsymbol{w}^{\ast },\boldsymbol{\theta })=g_{j}(\boldsymbol{w}%
			^{\ast },\boldsymbol{\theta }_{0})+\boldsymbol{h}_{j}(\boldsymbol{w}^{\ast },%
			\boldsymbol{\theta }_{0})(\boldsymbol{\theta -\theta }_{0})+\frac{1}{2}(%
			\boldsymbol{\theta -\theta }_{0})^{\prime }\boldsymbol{T}_{j}(\boldsymbol{w}%
			^{\ast },\boldsymbol{\bar{\theta}})(\boldsymbol{\theta -\theta }_{0}),
		\end{equation*}%
		where $\boldsymbol{\bar{\theta}}$ is a mean value, and therefore%
		\begin{align}
			\left\vert g_{j}(\boldsymbol{w}^{\ast },\boldsymbol{\theta })-g_{j}(%
			\boldsymbol{w}^{\ast },\boldsymbol{\theta }_{0})-\boldsymbol{h}_{j}(%
			\boldsymbol{w}^{\ast },\boldsymbol{\theta }_{0})(\boldsymbol{\theta -\theta }%
			_{0})\right\vert &=\frac{1}{2}\left\vert (\boldsymbol{\theta -\theta }%
			_{0})^{\prime }\boldsymbol{T}_{j}(\boldsymbol{w}^{\ast },\boldsymbol{\bar{%
					\theta}})(\boldsymbol{\theta -\theta }_{0})\right\vert  \notag \\
			&\leq \frac{1}{2}\left\Vert \boldsymbol{T}_{j}(\boldsymbol{w}^{\ast },%
			\boldsymbol{\bar{\theta}})\right\Vert \left\Vert \boldsymbol{\theta -\theta }%
			_{0}\right\Vert _{\infty }^{2},  \label{c bound_cf}
		\end{align}%
		by the Cauchy-Schwarz inequality. Define 
		\begin{equation*}
			\boldsymbol{a}^{\ast }\coloneqq%
			\begin{pmatrix}
				-\frac{\theta _{2}}{\theta _{1}^{2}} \\[4pt] 
				\frac{1}{\theta _{1}}%
			\end{pmatrix}%
			,
		\end{equation*}%
		and 
		\begin{equation*}
			\boldsymbol{B}^{\ast }\coloneqq%
			\begin{pmatrix}
				2\frac{\theta _{2}}{\theta _{1}^{3}} & -\frac{1}{\theta _{1}^{2}} \\[6pt] 
				-\frac{1}{\theta _{1}^{2}} & 0%
			\end{pmatrix}%
			.
		\end{equation*}
		Noting that the $\boldsymbol{h}_{j}(\boldsymbol{w}^{\ast },\boldsymbol{%
			\theta })$s are the $1\times 2$ rows of $\boldsymbol{H}_{\boldsymbol{\theta 
		}}(\boldsymbol{w}^{\ast },\boldsymbol{\theta })$, the componentwise bound (%
		\ref{c bound_cf}) implies 
		\begin{equation*}
			\left\Vert \boldsymbol{g}(\boldsymbol{w}^{\ast },\boldsymbol{\theta })-%
			\boldsymbol{g}(\boldsymbol{w}^{\ast },\boldsymbol{\theta }_{0})-\boldsymbol{H%
			}_{\boldsymbol{\theta }}(\boldsymbol{w}^{\ast },\boldsymbol{\theta }_{0})(%
			\boldsymbol{\theta -\theta }_{0})\right\Vert \leq \frac{1}{2}\left\Vert
			\left( \boldsymbol{T}_{1}(\boldsymbol{w}^{\ast },\boldsymbol{\bar{\theta}}%
			),\ldots ,\boldsymbol{T}_{k+2}(\boldsymbol{w}^{\ast },\boldsymbol{\bar{\theta%
			}})\right) \right\Vert \left\Vert \boldsymbol{\theta} -\boldsymbol{\theta} %
			_{0}\right\Vert _{\infty }^{2},
		\end{equation*}%
		where, for $j=1,\dots ,k+1$,%
		\begin{equation*}
			T_{j}(\boldsymbol{w}^{\ast },\boldsymbol{\theta })=-\,\phi \,q%
			_{j}^{\ast }(\zeta )%
			\begin{pmatrix}
				2\,\dfrac{\theta _{2}}{\theta _{1}^{3}} & -\dfrac{1}{\theta _{1}^{2}} \\%
				[6pt] 
				-\dfrac{1}{\theta _{1}^{2}} & 0%
			\end{pmatrix}%
			=-\phi q_{j}^{\ast }(\zeta )\boldsymbol{B}^{\ast },
		\end{equation*}%
		and%
		\begin{equation*}
			T_{k+2}(\boldsymbol{w}^{\ast },\boldsymbol{\theta })=%
			\begin{pmatrix}
				2\,\dfrac{\theta _{2}}{\theta _{1}^{3}}\,\varepsilon ^{\ast }\;-\;4\,\phi \,%
				\dfrac{\theta _{2}^{2}}{\theta _{1}^{4}} & -\,\dfrac{\varepsilon ^{\ast }}{%
					\theta _{1}^{2}}\;+\;3\,\phi \,\dfrac{\theta _{2}}{\theta _{1}^{3}} \\[8pt] 
				-\,\dfrac{\varepsilon ^{\ast }}{\theta _{1}^{2}}\;+\;3\,\phi \,\dfrac{\theta
					_{2}}{\theta _{1}^{3}} & -\,2\,\phi \,\dfrac{1}{\theta _{1}^{2}}%
			\end{pmatrix}%
			=\varepsilon ^{\ast }\boldsymbol{B}^{\ast }-\phi (\zeta \boldsymbol{B}^{\ast
			}+2\boldsymbol{a}^{\ast }\boldsymbol{a}^{\ast ^{\prime }}).
		\end{equation*}%
		Now, take 
		\begin{equation}
			\boldsymbol{l}(\boldsymbol{w}^{\ast },\boldsymbol{\theta })=\boldsymbol{H}_{%
				\boldsymbol{\theta }}(\boldsymbol{w}^{\ast },\boldsymbol{\theta }_{0})%
			\boldsymbol{\theta }  \label{l_2}
		\end{equation}%
		and $b(\boldsymbol{w}^{\ast })=\frac{1}{2}\left\Vert \left( \boldsymbol{T}%
		_{1}(\boldsymbol{w}^{\ast },\boldsymbol{\bar{\theta}}),\ldots ,\boldsymbol{T}%
		_{k+2}(\boldsymbol{w}^{\ast },\boldsymbol{\bar{\theta}})\right) \right\Vert $%
		. We have 
		\begin{align*}
			\mathbb{E}\left[b(\boldsymbol{w}^{\ast })\right] &=\mathbb{E}\left[ \vert\varepsilon
			^{\ast }\vert\left\Vert \boldsymbol{B}^{\ast }\right\Vert +\vert\phi \vert\left( \Vert
			\zeta \Vert \left\Vert \boldsymbol{B}^{\ast }\right\Vert +2\left\Vert 
			\boldsymbol{a}^{\ast }\right\Vert ^{2}\right)+\sum_{j=1}^{k+1}\vert {q%
			}_{j}^{\ast }(\zeta )\vert \, \vert\phi \vert\left\Vert \boldsymbol{B}^{\ast }\right\Vert
			\right] \\
			&=\mathbb{E}\left[\vert\varepsilon ^{\ast }\vert\right]\left\Vert \boldsymbol{B}^{\ast
			}\right\Vert +\vert\phi \vert\left( \Vert \zeta \Vert \left\Vert \boldsymbol{B}%
			^{\ast }\right\Vert +2\left\Vert \boldsymbol{a}^{\ast }\right\Vert ^{2}\right)
			+\sum_{j=1}^{k+1}\mathbb{E}\left[\vert q_{j}^{\ast }(\zeta )\vert\right]\vert\phi
			\vert\left\Vert \boldsymbol{B}^{\ast }\right\Vert .
		\end{align*}
		Since $z$ and $t$ are binary and by Assumption \ref{assumption moments Y}%
		, $\mathbb{E}[\vert\varepsilon ^{\ast 4}\vert]<\infty $ and also $%
		\mathbb{E}[\vert q_{j}^{\ast }(\zeta )\vert]<\infty $. Since, by
		Assumption \ref{assumption c}, $\boldsymbol{a}^{\ast }$ and $\boldsymbol{B}%
		^{\ast }$ are bounded, it follows that $\mathbb{E}\left[b(\boldsymbol{w}^{\ast
		})\right]<\infty $.
		
		\item[(b$^{*}$)] From (\ref{l_2}), $\left\Vert \boldsymbol{l}(\boldsymbol{w}%
		^{\ast },\boldsymbol{\theta })\right\Vert \leq \left\Vert \boldsymbol{H}_{%
			\boldsymbol{\theta }}(\boldsymbol{w}^{\ast },\boldsymbol{\theta }%
		_{0})\right\Vert \left\Vert \boldsymbol{\theta }\right\Vert _{\infty }$, so
		take $d(\boldsymbol{w}^{\ast })=\left\Vert \boldsymbol{H}_{\boldsymbol{%
				\theta }}(\boldsymbol{w}^{\ast },\boldsymbol{\theta }_{0})\right\Vert $. By
		Assumption \ref{assumption c}, which guarantees that $\theta _{1}$ is
		bounded way from zero, and by Assumption \ref{assumption moments Y} and the
		binary nature of $t$ and $z$, all entries of $\boldsymbol{H}_{\boldsymbol{%
				\theta }}(\boldsymbol{w}^{\ast },\boldsymbol{\theta }_{0})$ are bounded, and 
		$\mathbb{E}\left[d(\boldsymbol{w}^{\ast })^{2}\right]=\mathbb{E}\left[\left\Vert \boldsymbol{%
			H}_{\boldsymbol{\theta }}(\boldsymbol{w}^{\ast },\boldsymbol{\theta }%
		_{0})\right\Vert ^{2}\right]\leq \infty $.
		
		\item[(c$^{*}$)] From (\ref{l_2}), 
		\begin{eqnarray}
			\int \boldsymbol{l}(\boldsymbol{w}^{\ast },\boldsymbol{\theta })\,\mathrm{d}%
			F_{0}(\boldsymbol{w}^{\ast }) &=&\int \boldsymbol{H}_{\boldsymbol{\theta }}(%
			\boldsymbol{w}^{\ast },\boldsymbol{\theta }_{0})\boldsymbol{\theta }(%
			\boldsymbol{c})\,f_{0}(\boldsymbol{w}^{\ast })\,\mathrm{d}\boldsymbol{w}%
			^{\ast }  \notag \\
			&=&\int \boldsymbol{H}_{\boldsymbol{\theta }}(\boldsymbol{w}^{\ast },%
			\boldsymbol{\theta }_{0})f_{0}(\boldsymbol{w}^{\ast }\mid \boldsymbol{c})f_{0}(%
			\boldsymbol{c})\boldsymbol{\theta }(\boldsymbol{c})\,\mathrm{d}\boldsymbol{w}%
			^{\ast }  \notag \\
			&=&\int \mathbb{E}\left[ \boldsymbol{H}_{\boldsymbol{\theta }}(\boldsymbol{w}%
			^{\ast },\boldsymbol{\theta }_{0})\mid \boldsymbol{c}\right) f_{0}(\boldsymbol{c}%
			)\boldsymbol{\theta }(\boldsymbol{c})\,\mathrm{d}\boldsymbol{c}.
			\label{LHS_2}
		\end{eqnarray}%
		Equating (\ref{LHS_2}) to $\int \boldsymbol{v^{\ast }}(\boldsymbol{c})%
		\boldsymbol{\theta }(\boldsymbol{c})\,\mathrm{d}\boldsymbol{c}$, and using
		expression of $\boldsymbol{H}_{\boldsymbol{\theta }}(\boldsymbol{w}^{\ast
		},\theta _{o})$, we obtain 
		\begin{eqnarray}
			\underset{\left( k+2\right) \times 2}{\underbrace{\boldsymbol{v^{\ast }}(%
					\boldsymbol{c})}} &=&\mathbb{E}\left[ \boldsymbol{H}_{\boldsymbol{\theta }}(%
			\boldsymbol{w}^{\ast },\boldsymbol{\theta }_{0})\mid \boldsymbol{c}\right) f_{0}(%
			\boldsymbol{c})  \notag \\
			&=&\left( \mathbb{E}\left[ \varepsilon ^{\ast }\mid \boldsymbol{c}\right) 
			\boldsymbol{e}_{k+2,k+2}-\phi \,\mathbb{E}[\boldsymbol{q}_{0}^{\ast }\mid %
			\boldsymbol{c}]\right) \left( -\zeta _{0}(\boldsymbol{c}),1\right) .
			\label{v_2}
		\end{eqnarray}
		
		\item[(d$^{*}$)] The integral $\int \left\Vert \boldsymbol{v^{\ast }}(%
		\boldsymbol{c})\right\Vert \,\mathrm{d}\boldsymbol{c\ }$is finite because $%
		\boldsymbol{v}(\boldsymbol{c})$ is continuous almost everywhere by
		Assumption \ref{assumption smoothness}, is bounded on $\mathcal{C}$, and $%
		\mathcal{C}$ is compact. In particular, each conditional mean $\mathbb{E}[%
		q_{j}^{\ast }(\zeta )\mid \boldsymbol{c}]$ admits continuous,
		bounded versions on $\mathcal{C}$, and $\zeta (\boldsymbol{c})=\mathbb{E}[z\mid 
		\boldsymbol{c}]$ is bounded because $z$ is binary. Boundness of $\boldsymbol{%
			v^{\ast }}(\boldsymbol{c})$ also implies that there is $\epsilon >0$ such
		that $\mathbb{E}\left[\mathrm{sup}_{\left\Vert \boldsymbol{\mathrm{v}}\right\Vert \leq
			\epsilon }\left\Vert \boldsymbol{v}^{\ast }(\boldsymbol{c}+\boldsymbol{\mathrm{v}})\right\Vert
		^{4}\right]<\infty $.
		
		\item[(e$^{\ast}$)] Since $\mathbb{E}\left[\left\Vert \boldsymbol{g}(\boldsymbol{%
			w^{\ast }},\boldsymbol{\theta }_{0})\right\Vert ^{2}\right]=\mathbb{E}%
		(\sum_{j=1}^{k+2}g_{j}^{2}(\boldsymbol{w^{\ast }},\boldsymbol{\theta }_{0}))$%
		, it follows that $\mathbb{E}\left[\left\Vert \boldsymbol{g}(\boldsymbol{w^{\ast }%
		},\boldsymbol{\theta }_{0})\right\Vert ^{2}\right]<\infty $ if and only if $%
		\mathbb{E}\left[g_{j}^{2}(\boldsymbol{w^{\ast }},\boldsymbol{\theta }%
		_{0})\right]<\infty $ for each $j=1,\ldots ,k+2$. By the Cauchy-Schwarz inequality 
		$\mathbb{E}\left[g_{j}^{2}(\boldsymbol{w^{\ast }},\boldsymbol{\theta }_{0})\right]=%
		\mathbb{E}\left[q{_{j}^{2}}(\boldsymbol{\theta }_{0})(y-\boldsymbol{x}^{\ast
			\prime }\boldsymbol{\beta }_{0}^{\ast })^{2}\right]\leq \sqrt{\mathbb{E}\left[q{_{j}^{4}%
			}(\boldsymbol{\theta }_{0})\right]\mathbb{E}\left[(y-\boldsymbol{x}^{\ast \prime }%
			\boldsymbol{\beta }_{0}^{\ast })^{4}\right]},$ which, under Assumption \ref%
		{assumption moments Y}, is finite for each $j=2,\ldots ,k$. For $j=1$ note
		that $\mathbb{E}\left[q{_{1}^{4}}(\boldsymbol{\theta }_{0})\right]=\mathbb{E}\left[z{^{4}}%
		\right]<\infty $ and for $j=k+2$ we have $\mathbb{E}\left[q{_{k+2}^{4}}(\boldsymbol{%
			\theta }_{0})\right]=\mathbb{E}\left[\zeta _{0}^{4}(\boldsymbol{c})\right]<\infty $, due to 
		$z$ being binary.
	\end{enumerate}
	
	So far we have showed that Theorem 8.11 in NM applies to
	our setting. Theorem 8.12 in NM requires the following additional conditions.
	
	\begin{enumerate}
		\item[(f$^*$)] $\boldsymbol{\hat{\beta}}^{*}\xrightarrow[]{p}\boldsymbol{%
			\beta^* }_{0}$;
		
		\item[(g$^*$)] for $\Vert \boldsymbol{\theta }-\boldsymbol{\theta }_{0}\Vert
		_{\infty }$ sufficiently small, $\boldsymbol{g}(\boldsymbol{w}^{\ast },%
		\boldsymbol{\beta^* },\boldsymbol{\theta })$ is continuously differentiable
		in $\boldsymbol{\beta^* }$ on a neighborhood $\mathcal{N^*}$ of $\boldsymbol{%
			\beta^* }_{0}$, there exists a function $f(\boldsymbol{w}^{\ast })$ with $%
		\mathbb{E}\left[f(\boldsymbol{w}^{\ast })\right]<\infty $ and a constant $\varepsilon
		>0 $ such that $\Vert \boldsymbol{H}_{\boldsymbol{\beta^* }}(\boldsymbol{w}%
		^{\ast },\boldsymbol{\beta^* },\boldsymbol{\theta })-\boldsymbol{H}_{%
			\boldsymbol{\beta^* }}(\boldsymbol{w}^{\ast },\boldsymbol{\beta^* }_{0},%
		\boldsymbol{\theta }_{0})\Vert \leq f(\boldsymbol{w}^{\ast })\bigl(\Vert 
		\boldsymbol{\beta^* }-\boldsymbol{\beta^* }_{0}\Vert ^{\varepsilon }+\Vert 
		\boldsymbol{\theta }-\boldsymbol{\theta }_{0}\Vert _{\infty }^{\varepsilon }%
		\bigr)$;
		
		\item[(h$^*$)] $\mathbb{E}[\boldsymbol{H}_{\boldsymbol{\beta^* }}(%
		\boldsymbol{w}^{\ast },\boldsymbol{\theta }_{0})]$ exists and is nonsingular.
		
		Conditions (f$^*$)--(h$^*$) are verified next.
		
		\item[(f$^*$)] Under Assumptions \ref{assumption kernel}--\ref{assumption
			moments Y}, the Nadaraya-Watson estimator $\hat{\zeta}(\boldsymbol{c})$ is
		uniformly consistent. Adding Assumption \ref{assumption G*beta} , we also
		know that 
		\begin{equation*}
			\boldsymbol{\hat{\beta}}^{*}(\zeta _{0})\coloneqq\left( \sum_{i=1}^{n}%
			\boldsymbol{q}_{i}^*(\zeta _{0})\boldsymbol{x}_{i}^{*\prime }\right)
			^{-1}\left( \sum_{i=1}^{n}\boldsymbol{q}_{i}^*(\zeta _{0})y_{i}\right) %
			\xrightarrow[]{p}\boldsymbol{\beta^* }_{0}.
		\end{equation*}
		Consistency of the two-step estimator $\boldsymbol{\hat{\beta}}^{*}$ then
		follows from continuity of the map from $\hat{\zeta}$ to $\boldsymbol{\hat{%
				\beta}^*}$ and Slutsky's theorem.
		
		\item[(g$^*$)] From (\ref{Hbeta_cf}), 
		\begin{equation*}
			\boldsymbol{H}_{\boldsymbol{\beta^* }}(\boldsymbol{w}^{\ast },\boldsymbol{%
				\beta^* },\boldsymbol{\theta })-\boldsymbol{H}_{\boldsymbol{\beta^* }}(%
			\boldsymbol{w}^{\ast },\boldsymbol{\beta^* }_{0},\boldsymbol{\theta }%
			_{0})=-\left( \boldsymbol{q}(\boldsymbol{\theta })\boldsymbol{x}^{*\prime }(\boldsymbol{%
				\theta})-\boldsymbol{q}(\boldsymbol{\theta }_{0}) \boldsymbol{x}^{*\prime }(%
			\boldsymbol{\theta}_0)\right)=\boldsymbol{\Delta H_{\beta^*}}.
		\end{equation*}%
		We can rewrite this as, 
		\begin{equation*}
			\boldsymbol{\Delta H_{\beta^*}}=-\Bigl[(\boldsymbol{q}^*(\boldsymbol{\theta}%
			)-\boldsymbol{q}^*(\boldsymbol{\theta}_0))\,\boldsymbol{x}^{*\prime }(\boldsymbol{%
				\theta})+\boldsymbol{q}^*(\boldsymbol{\theta}_0)\,\bigl(%
			\boldsymbol{x}^{*\prime }(\boldsymbol{\theta})-\boldsymbol{x}^{*\prime }(\boldsymbol{%
				\theta}_0)\bigr)\Bigr] .
		\end{equation*}
		
		Now, if we take the norms and apply the triangle inequality we obtain 
		\begin{eqnarray*}
			\Vert\boldsymbol{\Delta H_{\beta*}}\Vert\leq \Vert\boldsymbol{q}^*(%
			\boldsymbol{\theta})-\boldsymbol{q}^*(\boldsymbol{\theta}_0)\Vert\Vert\,%
			\boldsymbol{x}^*(\boldsymbol{\theta})\Vert+ \Vert\boldsymbol{q}^*(%
			\boldsymbol{\theta}_0)\Vert \Vert \boldsymbol{x}^*(\boldsymbol{\theta})-%
			\boldsymbol{x}^*(\boldsymbol{\theta}_0)\Vert.
		\end{eqnarray*}
		
		We now show that, for $\Vert \boldsymbol{\theta }-\boldsymbol{\theta }%
		_{0}\Vert _{\infty }$ sufficiently small, $\Vert \boldsymbol{\Delta
			H_{\beta^*}}\Vert \leq f(\boldsymbol{w^*})\Vert \boldsymbol{\theta }-%
		\boldsymbol{\theta }_{0}\Vert _{\infty }$. Since $\boldsymbol{q^*}(%
		\boldsymbol{\theta })-\boldsymbol{q^*}(\boldsymbol{\theta }_{0})=(%
		\boldsymbol{0}_{k+1}^{\prime },\zeta -\zeta _{0})^{\prime }$,%
		\begin{eqnarray}
			\left\Vert \boldsymbol{q^*}(\boldsymbol{\theta })-\boldsymbol{q^*}(%
			\boldsymbol{\theta }_{0})\right\Vert &=&\left\vert \zeta -\zeta
			_{0}\right\vert  \notag \\
			&=&\left\vert \frac{\theta _{2}\theta _{1,0}-\theta _{2,0}\theta _{1}}{%
				\theta _{1}\theta _{1,0}}\right\vert .  \label{ratio_2}
		\end{eqnarray}
		
		The numerator of (\ref{ratio_2}) is $\theta _{2}\theta _{1,0}-\theta
		_{2,0}\theta _{1}=(\theta _{2}-\theta _{2,0})\theta _{1,0}+\theta
		_{2,0}(\theta _{1,0}-\theta _{1})$, so by the triangle inequality $%
		\left\vert \theta _{2}\theta _{1,0}-\theta _{2,0}\theta _{1}\right\vert \leq
		\left\vert (\theta _{2}-\theta _{2,0})\theta _{1,0}\right\vert +\left\vert
		\theta _{2,0}(\theta _{1,0}-\theta _{1})\right\vert \leq \left\vert (\theta
		_{2}-\theta _{2,0})\theta _{1,0}\right\vert +\left\vert \theta _{2,0}(\theta
		_{1,0}-\theta _{1})\right\vert .$ By Assumptions \ref{assumption c} and \ref{assumption smoothness}, $\left\vert \theta _{2}\theta _{1,0}-\theta
		_{2,0}\theta _{1}\right\vert \leq C\Vert \boldsymbol{\theta }-\boldsymbol{%
			\theta }_{0}\Vert _{\infty }$, where $C$ is a constant that depends on the
		smoothness of $\mathbb{E}\left[z\mid \boldsymbol{c}\right]$ and the bound on $f(\boldsymbol{%
			c})$. As for the denominator of (\ref{ratio_2}), by Assumption \ref%
		{assumption c}, there exists a constant $\delta >0$ such that $\theta
		_{1}\theta _{1,0}\geq \delta $ for all $\boldsymbol{c}\in \mathcal{C}.$
		Combining the bounds on the numerator and denominator, we obtain 
		\begin{equation}
			\left\vert
			\zeta (\boldsymbol{c})-\zeta _{0}(\boldsymbol{c})\right\vert \leq \delta
			^{-1}C\Vert \boldsymbol{\theta }-\boldsymbol{\theta }_{0}\Vert _{\infty }.\label{delta2}
		\end{equation}
		
		Note that since $\boldsymbol{x^*}(\boldsymbol{\theta })-\boldsymbol{x^*}(%
		\boldsymbol{\theta }_{0})=(\boldsymbol{0}_{k+1}^{\prime }, \zeta -\zeta
		_{0})^{\prime }$,%
		\begin{eqnarray}
			\left\Vert \boldsymbol{x^*}(\boldsymbol{\theta })-\boldsymbol{x^*}(%
			\boldsymbol{\theta }_{0})\right\Vert &=&\left\vert \zeta -\zeta
			_{0}\right\vert  \notag \\
			&=&\left\vert \frac{\theta _{2}\theta _{1,0}-\theta _{2,0}\theta _{1}}{%
				\theta _{1}\theta _{1,0}}\right\vert ,  \label{ratio2_2}
		\end{eqnarray}
		and so the same exact arguments hold. Also note that in a ball $\{\boldsymbol{\theta}:\Vert 
		\boldsymbol{\theta}-\boldsymbol{\theta}_0\Vert \leq\mathbb{\epsilon}\}$,
		continuity of $\boldsymbol{x}^*(\boldsymbol{\theta})$ gives $\Vert 
		\boldsymbol{x}^*(\boldsymbol{\theta})\Vert \leq 2\Vert \boldsymbol{x}^*(%
		\boldsymbol{\theta}_0)\Vert$. Hence, in this case, take $f(\boldsymbol{w^*})=2\left \Vert \boldsymbol{x^*(\theta_0)}\right \Vert$+ $\left\Vert 
		\boldsymbol{q{^*}(\theta_0)}\right\Vert$.\footnote{%
			Pick $\mathbb{\epsilon}>0$ small enough that whenever $\|\boldsymbol{\theta} -
			\boldsymbol{\theta}_0\|_\infty \le \mathbb{\epsilon}$, we have $\Vert \boldsymbol{x}^*(\boldsymbol{\theta}) -
			\boldsymbol{x}^*(\boldsymbol{\theta}_0)\Vert \le \Vert \boldsymbol{x}^*(\boldsymbol{\theta}_0)\Vert.$ Then by the triangle
			inequality, $
				\Vert \boldsymbol{x}^*(\boldsymbol{\theta})\Vert \;\le\; \Vert \boldsymbol{x}^*(\boldsymbol{\theta}_0)\Vert \;+\; \Vert
				\boldsymbol{x}^*(\boldsymbol{\theta}) - \boldsymbol{x}^*(\boldsymbol{\theta}_0)\Vert \;\le\; \Vert \boldsymbol{x}^*(\boldsymbol{\theta}_0)\Vert \;+\;
				\Vert \boldsymbol{x}^*(\boldsymbol{\theta}_0)\Vert \;=\; 2\,\Vert \boldsymbol{x}^*(\boldsymbol{\theta}_0)\Vert.$	} Then $\mathbb{E}\left[f(\boldsymbol{w^*})\right]<\infty $ by Assumption \ref%
		{assumption moments Y} and the fact that $t$ and $z$ are binary.
		
		\item[(h$^{*}$)] Existence is guaranteed by Assumption \ref{assumption
			moments Y} and, by the binary nature of $z$, nonsingularity by Assumption \ref%
		{assumption G*beta}.
	\end{enumerate}
	
	Having showed that Theorems 8.11 and 8.12 in NM apply to our context, to
	complete the proof we only need to derive the expression for the $\left(
	k+2\right) \times 1$ correction term $\boldsymbol{\varphi }(\boldsymbol{w}%
	^{\ast })$. From (\ref{v_2}),%
	\begin{eqnarray*}
		\boldsymbol{v^{\ast }}(\boldsymbol{c})\boldsymbol{\tilde{z}} &=&\big( 
		\mathbb{E}\left[ \varepsilon ^{\ast }\mid\boldsymbol{c}\right] \boldsymbol{e}%
		_{k+2,k+2}-\phi \mathbb{E}\left[\boldsymbol{q}_{0}^{\ast }\mid\boldsymbol{c}\right]\big)
		\left( -\zeta _{0}(\boldsymbol{c}),1\right) 
		\begin{pmatrix}
			1 \\ 
			z%
		\end{pmatrix}
		\\
		&=&\big( \mathbb{E}\left[ \varepsilon ^{\ast }\mid\boldsymbol{c}\right] 
		\boldsymbol{e}_{k+2,k+2}-\phi \mathbb{E}\left[ \boldsymbol{q}_{0}^{\ast }\mid%
		\boldsymbol{c}\right] \big) z^{\ast }(\zeta _{0}),
	\end{eqnarray*}%
	where 
	\begin{equation*}
		\mathbb{E}\left[\boldsymbol{q}_{0}^{\ast }\mid\boldsymbol{c}\right]=%
		\begin{pmatrix}
			\zeta _{0}(\boldsymbol{c}) \\ 
			\boldsymbol{r}_{k}^{\prime } \\ 
			\zeta _{0}(\boldsymbol{c})%
		\end{pmatrix}%
		,
	\end{equation*}%
	and therefore $\boldsymbol{\varphi }(\boldsymbol{w^{\ast }})$ is a $(k+2)\times 1$
	vector with entries 
	\begin{eqnarray*}
		\varphi _{1}(\boldsymbol{w^{\ast }}) &=&-\phi \zeta _{0}(\boldsymbol{c}%
		)z^{\ast }(\zeta _{0})-\mathbb{E}\left[ -\phi \zeta _{0}(\boldsymbol{c}%
		)z^{\ast }(\zeta _{0})\right]  \\
		&=&-\phi \zeta _{0}(\boldsymbol{c})z^{\ast }(\zeta _{0}),
	\end{eqnarray*}
	\vspace{-1.2cm}
	\begin{eqnarray*}
		\varphi _{j}(\boldsymbol{w^{\ast }}) &=&-\phi \boldsymbol{r}_{j-1}z^{\ast
		}(\zeta _{0})-\mathbb{E}\left[ -\phi \boldsymbol{r}_{j-1}z^{\ast }(\zeta
		_{0})\right]  \\
		&=&-\phi \boldsymbol{r}_{j-1}z^{\ast }(\zeta _{0}),
	\end{eqnarray*}%
	for $j=2,...,k+1$, and 
	\begin{eqnarray*}
		\varphi _{k+2}(\boldsymbol{w^{\ast }}) &=&\big[ \mathbb{E}\left[
		\varepsilon ^{\ast }\mid\boldsymbol{c}\right] -\phi \zeta _{0}(\boldsymbol{c})%
		\big] z^{\ast }(\zeta _{0})-\mathbb{E}\big[ \left( \mathbb{E}\left[
		\varepsilon ^{\ast }\mid\boldsymbol{c}\right] -\phi \zeta _{0}(\boldsymbol{c})%
		\right) z^{\ast }(\zeta _{0})\big]  \\
		&=&\big[ \mathbb{E}\left[ \varepsilon ^{\ast }\mid\boldsymbol{c}\right] -\phi
		\zeta _{0}(\boldsymbol{c})\big] z^{\ast }(\zeta _{0}),
	\end{eqnarray*}%
	since $\mathbb{E}\big[ -\phi \mathbb{E}\left[ \boldsymbol{q}_{0}^{\ast }\mid%
	\boldsymbol{c}\right] \left( z-\mathbb{E}\left[ z\mid\boldsymbol{c}\right] %
	\right) \big] =0$.
\end{pff}

\begin{pff}[Proof of Theorem \protect\ref{cons var CF}]
	The result follows from Theorem 8.13 in NM, which requires a regularity condition that can be broken down in the two following conditions:
	
	\begin{itemize}
		\item[(i$^*$)] there exists a function $b(\boldsymbol{w^*})$ and constant $\epsilon > 0$ such that $\mathbb{E}[b(\boldsymbol{w^*})^2] < \infty$, and for $\Vert \boldsymbol{\theta} - \boldsymbol{\theta}_0 \Vert_\infty$ small enough,
		$\Vert \boldsymbol{g}(\boldsymbol{w}^*, \boldsymbol{\beta}^*, \boldsymbol{\theta}) - \boldsymbol{g}(\boldsymbol{w}^*, \boldsymbol{\beta}^*_0, \boldsymbol{\theta}_0)\Vert 
		\leq b(\boldsymbol{w}^*) \left( \Vert\boldsymbol{\beta}^* - \boldsymbol{\beta}^*_0\Vert^\epsilon + \Vert\boldsymbol{\theta} - \boldsymbol{\theta}_0\Vert_\infty^\epsilon \right);$
		
		\item[(j$^*$)] there exists a function $d(\boldsymbol{w}^*)$ and constant $\epsilon > 0$ such that $\mathbb{E}[d(\boldsymbol{w}^*)^2] < \infty$, and for $\Vert \boldsymbol{\theta} - \boldsymbol{\theta}_0 \Vert_\infty$ small enough, $
		\Vert\boldsymbol{\varphi}(\boldsymbol{w}^*, \boldsymbol{\beta}^*, \boldsymbol{\theta}) - \boldsymbol{\varphi}(\boldsymbol{w}^*, \boldsymbol{\beta}^*_0, \boldsymbol{\theta}_0)\Vert
		\leq d(\boldsymbol{w}^*) \left( \Vert\boldsymbol{\beta}^* - \boldsymbol{\beta}^*_0\Vert^\epsilon + \Vert\boldsymbol{\theta} - \boldsymbol{\theta}_0\Vert_\infty^\epsilon \right).$

	\end{itemize}
	
	Conditions ($i^*$) and ($j^*$) are verified as follows.
	
	\begin{itemize}
		\item[(i$^*$)] Note that
		\begin{align*}
			\Delta \boldsymbol g(\boldsymbol w^*) &\coloneqq \boldsymbol g(\boldsymbol{w}^*, \boldsymbol{\beta}^*, \boldsymbol{\theta}) - \boldsymbol  g(\boldsymbol{w}^*, \boldsymbol{\beta}_0^*, \boldsymbol{\theta}_0) \\
			&= \boldsymbol{q}^*(\boldsymbol{\theta}) \boldsymbol{x}^*(\boldsymbol{\theta})^\prime (\boldsymbol{\beta}_0^* - \boldsymbol{\beta}^*) 
			+ (y - \boldsymbol{x}^*(\boldsymbol{\theta})^\prime \boldsymbol{\beta}_0^*) (\boldsymbol{q}^*(\boldsymbol{\theta}) - \boldsymbol{q}^*(\boldsymbol{\theta}_0)) \\
			&\quad + \boldsymbol{q}^*(\boldsymbol{\theta}_0)(\boldsymbol{x}^*(\boldsymbol{\theta}_0)^\prime - \boldsymbol{x}^*(\boldsymbol{\theta})^\prime) \boldsymbol{\beta}_0^*.
		\end{align*}
		
		By the triangle inequality,
		\begin{align*}
			\Vert\Delta \boldsymbol g(\boldsymbol{w}^*)\Vert &\leq \Vert\boldsymbol{q}^*(\boldsymbol{\theta})\Vert  \Vert\boldsymbol{x}^*(\boldsymbol{\theta})\Vert  \Vert\boldsymbol{\beta}^* - \boldsymbol{\beta}_0^*\Vert \\
			&\quad + \vert y - \boldsymbol{x}^*(\boldsymbol{\theta})^\prime \boldsymbol{\beta}_0^*\vert  \Vert\boldsymbol{q}^*(\boldsymbol{\theta}) - \boldsymbol{q}^*(\boldsymbol{\theta}_0)\Vert \\
			&\quad + \Vert\boldsymbol{q}^*(\boldsymbol{\theta}_0)\Vert  \Vert\boldsymbol{\beta}_0^*\Vert  \Vert\boldsymbol{x}^*(\boldsymbol{\theta}) - \boldsymbol{x}^*(\boldsymbol{\theta}_0)\Vert.
		\end{align*}
		
		Since $\boldsymbol{q}^*(\boldsymbol{\theta})$ and $\boldsymbol{x}^*(\boldsymbol{\theta})$ depend on $\zeta(\boldsymbol{c})$, which is Lipschitz in $\boldsymbol{\theta}$, as shown in equations (\ref{ratio_2}) and (\ref{ratio2_2}), 
		we can write
		\[
		b(\boldsymbol{w}^*) \coloneqq   \max \Big\{ 
		\sup_{\theta} \Vert\boldsymbol{q}^*(\theta)\Vert \Vert\boldsymbol{x}^*(\theta)\Vert,\sup_{\theta} \delta^{-1}C \big(
		\vert y - \boldsymbol{x}^*(\theta)^\prime \boldsymbol{\beta}_0^*\vert+\ 
		\Vert\boldsymbol{q}^*(\theta_0)\Vert \Vert\boldsymbol{\beta}_0^*\Vert\big)
		\Big\},
		\]
		which is bounded by Assumption \ref{assumption moments Y} and by the binary nature of $t$ and $z$. We conclude that $\mathbb{E}[b(\boldsymbol{w^*})^2] < \infty$.
		\bigskip
		
		\item[(j$^*$)] Recall that $\boldsymbol{\varphi}(\boldsymbol{w}^*)$ is a $(k+2)$ vector. 
		For $j=1$,
		\[
		\begin{aligned}
			\varphi_1(\boldsymbol w^*, \boldsymbol \beta^*, \boldsymbol \theta) - \varphi_1(\boldsymbol w^*, \boldsymbol \beta_0^*, \boldsymbol \theta_0)
			&= -\phi\, \zeta(\boldsymbol c)\, z^*(\zeta) + \phi_0\, \zeta_0(\boldsymbol c)\, z^*(\zeta_0) \\
			&= -\left[ \phi\, \zeta(\boldsymbol c) - \phi_0\, \zeta_0(\boldsymbol c) \right] z^*(\zeta) \\
			&\quad - \phi_0 \zeta_0(\boldsymbol c) \left[ \zeta_0(\boldsymbol c) - \zeta(\boldsymbol c) \right].
		\end{aligned}
		\]
		By the triangle inequality
		\[
		\begin{aligned}
			\left\Vert \varphi_1(\boldsymbol w^*, \boldsymbol \beta^*, \boldsymbol \theta) - \varphi_1(\boldsymbol w^*, \boldsymbol \beta_0^*, \boldsymbol \theta_0) \right\Vert
			&\leq \left\vert \phi\, \zeta(\boldsymbol c) - \phi_0\, \zeta_0(\boldsymbol c) \right\vert \, \left\vert z^*(\zeta) \right\vert \\
			&\quad + \left\vert \phi_0\, \zeta_0(\boldsymbol c) \right\vert \, \left\vert \zeta(\boldsymbol c) - \zeta_0(\boldsymbol c) \right\vert \\
			&\leq \big( \left\vert \phi - \phi_0 \right\vert \, \left\vert \zeta(\boldsymbol c) \right\vert
			+ \left\vert \phi_0 \right\vert  \, \left\vert \zeta(\boldsymbol c) - \zeta_0(\boldsymbol c) \right\vert \big)  \left\vert z^*(\zeta) \right\vert \\
			&\quad + \left\vert \phi_0\, \zeta_0(\boldsymbol c) \right\vert \, \left\vert \zeta(\boldsymbol c) - \zeta_0(\boldsymbol c) \right\vert.
		\end{aligned}
		\]
		Using equation (\ref{delta2}),
		define
		\[
		d_1(\boldsymbol w^*) \coloneqq \max \left\{
		\sup_{\theta} \delta^{-1} C  \big( \vert \phi_0 \vert \, \vert z^*(\zeta) \vert + \vert \phi_0 \zeta_0(\boldsymbol c) \vert \big),
		\sup_{\theta } \vert \zeta(\boldsymbol c) \vert
		\right\}.
		\]
		
		For $j = 2, \dots, k+1$ we have
		\[
		\begin{aligned}
			\varphi_j(\boldsymbol w^*, \boldsymbol \beta^*, \boldsymbol \theta) - \varphi_j(\boldsymbol w^*, \boldsymbol \beta_0^*, \boldsymbol \theta_0)
			&= -(\phi - \phi_0) \boldsymbol r_{j-1}z^*(\zeta) - \phi_0 \boldsymbol r_{j-1} (\zeta(\boldsymbol c) - \zeta_0(\boldsymbol c)).
		\end{aligned}
		\]
		Applying the triangle inequality gives
		\[
		\left\Vert \varphi_j(\boldsymbol w^*, \boldsymbol \beta^*, \boldsymbol \theta) - \varphi_j(\boldsymbol w^*, \boldsymbol \beta_0^*, \boldsymbol \theta_0) \right\Vert 
		\leq \vert \phi - \phi_0 \vert  \Vert \boldsymbol r_{j-1} \Vert  \vert z^*(\zeta) \vert
		+ \vert \phi_0 \vert  \vert \zeta(\boldsymbol c) - \zeta_0(\boldsymbol c) \vert  \Vert \boldsymbol r_{j-1} \Vert.
		\]
		Using again (\ref{delta2}),
		\[
		d_2(\boldsymbol w^*) \coloneqq \max \left\{
		\delta^{-1} C  \vert\phi_0\vert  \Vert \boldsymbol r_{j-1} \Vert,
		\sup_{\theta} \vert z^*(\zeta) \vert  \Vert \boldsymbol r_{j-1} \Vert
		\right\}.
		\]
		
		Finally, for $j = k+2$, start by defining $\mu_y(\boldsymbol c) \coloneqq \mathbb{E}\left[y \mid \boldsymbol c\right], \text{and } 
		\boldsymbol \mu_x^*(\boldsymbol c; \boldsymbol \theta) \coloneqq \mathbb{E}\left[\boldsymbol x^*(\boldsymbol \theta) \mid \boldsymbol c\right].$
		Therefore we can write
		\[
		\varphi_{k+2}(\boldsymbol w^*, \boldsymbol \beta^*, \boldsymbol \theta)
		\coloneqq \left[ \mu_y(\boldsymbol c) - \boldsymbol \mu_x^*(\boldsymbol c; \boldsymbol \theta)' \boldsymbol \beta^* - \phi \zeta(\boldsymbol c) \right] z^*(\zeta),
		\]
		and
		\[
		\begin{aligned}
			&\varphi_{k+2}(\boldsymbol w^*, \boldsymbol \beta^*, \boldsymbol \theta) - \varphi_{k+2}(\boldsymbol w^*, \boldsymbol \beta_0^*, \boldsymbol \theta_0) \\
			&= \left[
			\boldsymbol \mu_x^*(\boldsymbol c; \boldsymbol \theta_0)' \boldsymbol \beta_0^*
			- \boldsymbol \mu_x^*(\boldsymbol c; \boldsymbol \theta)' \boldsymbol \beta^*
			+ \phi_0 \zeta_0(\boldsymbol c) - \phi \zeta(\boldsymbol c)
			\right] z^*(\zeta) \\
			&\quad + \left[
			\mu_y(\boldsymbol c) - \boldsymbol \mu_x^*(\boldsymbol c; \boldsymbol \theta_0)' \boldsymbol \beta_0^* - \phi_0 \zeta_0(\boldsymbol c)
			\right] \left[ z^*(\zeta) - z^*(\zeta_0) \right].
		\end{aligned}
		\]
		
		Now applying the triangle inequality
		\[
		\begin{aligned}
			\left\Vert \varphi_{k+2}(\boldsymbol w^*, \boldsymbol \beta^*,\boldsymbol \theta) - \varphi_{k+2}(\boldsymbol w^*, \boldsymbol \beta_0^* ,\boldsymbol \theta_0) \right\Vert
			&\leq \left\Vert \boldsymbol \mu_x^*(\boldsymbol c; \boldsymbol \theta) - \boldsymbol \mu_x^*(\boldsymbol c; \boldsymbol \theta_0) \right\Vert  \left\Vert \boldsymbol \beta^* \right\Vert  \left\vert z^*(\zeta) \right\vert \\
			&\quad + \left\Vert \boldsymbol \mu_x^*(\boldsymbol c; \boldsymbol \theta_0) \right\Vert  \left\Vert \boldsymbol \beta^* - \boldsymbol \beta_0^* \right\Vert  \left\vert z^*(\zeta) \right\vert \\
			&\quad + \left\vert \phi \zeta(\boldsymbol c) - \phi \zeta_0(\boldsymbol c) \right\vert \, \left\vert z^*(\zeta) \right\vert \\
			&\quad + \left\vert \mu_y(\boldsymbol c) - \boldsymbol \mu_x^*(\boldsymbol c; \boldsymbol \theta_0)' \boldsymbol \beta_0^* - \phi_0 \zeta_0(\boldsymbol c) \right\vert \, \left\vert z^*(\zeta) - z^*(\zeta_0) \right\vert.
		\end{aligned}
		\]
		Since $\boldsymbol \mu_x^*(\boldsymbol c; \boldsymbol \theta) - \boldsymbol \mu_x^*(\boldsymbol c; \boldsymbol \theta_0) = (\boldsymbol 0_{k+1}, \zeta_0 - \zeta)$, and using (\ref{ratio_2}), define
		\[
		\begin{aligned}
			\begin{aligned}
				d_3(\boldsymbol w^*) \coloneqq \max  \Big\{\sup_{\theta}(
				&\Vert \boldsymbol \beta_0^* \Vert  \vert z^*(\zeta) \vert + \vert z^*(\zeta) \vert + \left\vert \mu_y(\boldsymbol c) - \boldsymbol \mu_x^*(\boldsymbol c; \boldsymbol \theta_0)' \boldsymbol \beta_0^* - \phi_0 \zeta_0(\boldsymbol c) \right\vert), \\
				&\sup_{\theta}\Vert \boldsymbol \mu_x^*(\boldsymbol c; \boldsymbol \theta_0) \Vert  \vert z^*(\zeta) \vert \Big\}.
			\end{aligned}
		\end{aligned}
		\]
		
		Note that $d_1(\boldsymbol w^*), d_2(\boldsymbol w^*) \text{ and } d_3(\boldsymbol w^*)$ are finite given Assumptions \ref{assumption c} and \ref{assumption moments Y}, along with the bounded support and binary structure of $z \text{ and } t$. Therefore, to conclude, we can take $d(\boldsymbol w^*)$ to be $\max \left\{ d_1(\boldsymbol w^*), d_2(\boldsymbol w^*), d_3(\boldsymbol w^*) \right\}$, which clearly satisfies $
		\mathbb{E}[d(\boldsymbol w^*)^2] < \infty.$
	\end{itemize}
	
\end{pff}

\end{appendices}

\bibliographystyle{ecta}
\bibliography{referencesLATE}

\end{document}